\documentclass[preprint,12pt]{elsarticle}

\usepackage{amssymb}
\usepackage{amsthm}

\usepackage{lineno}
\usepackage{graphicx}
\usepackage{dcolumn}
\usepackage{bm}
\usepackage{rotating,booktabs}
\usepackage{threeparttable}

\journal{Nuclear Physics A}

\providecommand{\Journal}[4]{#1{\bf #2} (#4) #3}
\providecommand{\EPJA}{Eur. Phys. J. A} %
\providecommand{\NPA}{Nucl. Phys. A} %
\providecommand{\NPB}{Nucl. Phys. B} %
\providecommand{\PL}{Phys. Lett. } %
\providecommand{\PLB}{Phys. Lett. B } %
\providecommand{\PR}{Phys. Rev. } %
\providecommand{\PRL}{Phys. Rev. Lett. } %
\providecommand{\PRC}{Phys. Rev. C } %
\providecommand{\PRD}{Phys. Rev. D } %
\providecommand{\PRN}{Prog. Part. Nucl. Phys. } %
\providecommand{\Science}{Science } %
\providecommand{\Ann}{ Ann.Rev.Nucl.Part.Sci.}
\begin{document}

\begin{frontmatter}



\title{Light flavor baryon spectrum with higher order hyperfine interactions}

\author{Yan Chen}
\author{Bo-Qiang Ma
}\ead{mabq@phy.pku.edu.cn}

\address{School of Physics and State Key Laboratory of Nuclear
Physics and Technology, Peking University, Beijing 100871, China}

\begin{abstract}
We study the spectrum of light flavor baryons in a quark-model
framework  by taking into account the order $\mathrm{O}(\alpha_s^2)$
hyperfine interactions due to two-gluon exchange between quarks. The
calculated spectrum agree better with the experimental data than the
results from hyperfine interactions with only one-gluon exchange. It
is also shown that two-gluon exchange hyperfine interactions bring a
significantly improved correction to the Gell-Mann--Okubo mass
formula. Two-gluon exchange corrections on baryon excitations
(including negative parity baryons) are also briefly discussed.

\end{abstract}


\end{frontmatter}

\newpage


\section{Introduction}
A number of approaches have been developed for describing the baryon
spectrum since the 1960's, such as the SU(6)
model~\cite{Gursey,Sakita}, the quark
model~\cite{RujulaCapstick,Capstick}, the bag
model~\cite{Chodosa,Chodosb,DeGrand}, the Skyrme
model~\cite{Hayasgi140,M.P.Mattis}, the large $N_c$ baryon
model~\cite{Hooft,witten,Jenkins,Carlson,Matagne,Goity,Matagne2005}
and so on. These models incorporate partly the dynamics of quantum
chromodynamics (QCD) and have arrived at remarkable (though not
quantitatively perfect) success. Recent lattice calculations have
made significant progress on the spectrum of bound states and offer
a bright promise~\cite{Allton,Aoki,Aubin,Jansen,Durr}. However, the
quark model still remains a basic and indispensable tool for
understanding hadron spectroscopy due to its intuition and
simplicity as a guide-line to other approaches. The quark model has
been proved quite fruitful on the study of baryon spectrum, decays
and moments. In the quark model with dynamics, the spectrum of
hadrons is dominated by two ingredients: the long range forces to
bind quarks as hadrons and the short range forces expected from
gluon exchanges. The long range forces are the confining forces,
which are flavor and spin independent. The short range forces
include a Coulomb term, hyperfine interaction including the
spin-orbit interaction. Actually spin-spin part of the hyperfine
interaction is
 the most important short range force that is responsible for such prominent features such as
the $\Delta$-$N$ and $\rho$-$\pi$ mass
splittings~\cite{RujulaCapstick}.

The hyperfine interaction employed in quark model is usually derived
from one-gluon exchange and it contains a spin-spin and a tensor
part. In baryons the spin-orbit interaction can be neglected. The
next step for a further improvement of the hyperfine interaction is
to consider not only one-gluon exchange but also higher order terms
such as two-gluon exchange. In fact, such a potential with two-gluon
exchange has already been calculated by Gupta and
Radford~\cite{Gupta and Radford24,Gupta and Radford25}. Quarkonium
spectra were investigated with the higher order
potential~\cite{Gupta and Radford26,Titard94,J.Eiglsperger} and the
theoretical results agree excellently with experiments. Inspired by
these results, we try to investigate the light flavor baryon
spectrum with the hyperfine interaction of higher order
$\mathrm{O}(\alpha_s^2)$. Since the nonrelativistic quark model is
the simplest and most economical quark potential model with
considerable phenomenological success~\cite{Isgur and Karl77,Isgur
and Karl78,Isgur and Karl4187,Isgur and Karl2653,Isgur and
Karl1191,Kuang-Ta Chao and Isgur 155,Buchmann97}, we try to study
the nonrelativistic quark model with the hyperfine interaction of
higher order $\mathrm{O}(\alpha_s^2)$. Furthermore, we particularly
focus our attention on the ground-state baryon masses, as all wave
functions of ground-state baryons have zero orbital angular momentum
so that we can neglect the spin-orbit forces which should be better
to be treated in a relativistic framework.

In this paper, we adopt the simple harmonic-oscillator quark model
with the interactions between quarks being harmonic springs and
hyperfine interactions. The hyperfine interaction includes not only
one-gluon exchange of order $\alpha_s$, but also two-gluon exchange
of order $\alpha_s^2$ . After choosing the traditional zero-order
wave functions of the SU(6) multiplets $(56,0^+)$, $(70,1^-)$,
$(56',0^+)$, $(70,0^+)$, $(56,2^+)$, $(70,2^+)$ and $(20,1^+)$ at
$N=0$, $N=1$ and $N=2$ respectively, we employ the Hamiltonian and
the baryon wave functions to calculate the matrix elements. Since
the hyperfine matrix elements of $\mathrm{O}(\alpha_s)$  have been
calculated in previous studies, we only calculate the hyperfine
matrix elements of
 $\mathrm{O}(\alpha_s^2)$. We then diagonalize the complete matrices to get
the  baryon masses. We find that the masses are mostly in agreement
with the experimental masses. In addition, the masses of the
ground-state baryons are better than those with only one-gluon
exchange. We also find that the two-gluon exchange interactions
bring a significant improved correction to the Gell-Mann--Okubo mass
formula of the ground-state baryons. We also briefly discuss the
two-gluon exchange corrections on negative-parity baryons and other
baryon excitations.

\section{\label{sec:2} The Hamiltonian and its solutions}
 The Hamiltonian in the quark model~\cite{Isgur and
Karl2653} is
\begin{eqnarray}
H&=&\sum_i m_i +H_0+H_{\mathrm{hyp}},
\end{eqnarray}
\begin{eqnarray}
H_0&=&\sum_i \frac{p_i^2}{2m_i}+\sum_{i<j} V_{\mathrm{conf}}^{ij},
\end{eqnarray}
\begin{eqnarray}
H_{\mathrm{hyp}}&=&\sum_{i<j}H_{\mathrm{hyp}}^{ij},
\end{eqnarray}
where $V_{\mathrm{conf}}^{ij}$ is the spin-independent potential and
$H_{\mathrm{hyp}}^{ij}$ is the hyperfine interaction. The potential
$V_{\mathrm{conf}}^{ij}$ is
\begin{eqnarray}
V_{\mathrm{conf}}^{ij}=\frac{1}{2}Kr_{ij}^2+U(r_{ij}),
\end{eqnarray}
where anharmonicity $U(r_{ij})$ is a function which only depends on
$r_{ij}$, and the explicit form of $U(r_{ij})$ is not needed in this
paper.

The  hyperfine interaction $H_{\mathrm{hyp}}^{ij}$ is
\begin{eqnarray}
H_{\mathrm{hyp}}^{ij}&=&H_{\mathrm{hyp}}^{ij}(\alpha_s)+H_{\mathrm{hyp}}^{ij}(\alpha_s^2),
\end{eqnarray}
where we add the order $\alpha_s^2$  interaction.

The hyperfine interaction of  order $\alpha_s$  derived from the
one-gluon exchange process is
\begin{eqnarray}
H_{\mathrm{hyp}}^{ij}(\alpha_s)&=&\frac{2
\alpha_s}{3m_im_j}[\frac{8\pi}{3}{\vec{S}_i}\cdot
{\vec{S}_j}\delta^3(\vec{r}_{ij})\nonumber\\
&&+\frac{1}{r_{ij}^3}(\frac{3{\vec{S}_i}\cdot \vec{r}_{ij}\
{\vec{S}_j}
\cdot\vec{r}_{ij}}{r_{ij}^2}-{\vec{S}_i}\cdot{\vec{S}_j})],
\end{eqnarray}
where $m_i$ and ${\vec{S}_i}$ are the mass and spin of the $i$th
quark and $\vec{r}_{ij}$ is the relative position of the pair $ij$
of quarks. The first term is called the Fermi contact term and
operates when the quark pair has zero orbital angular momentum. The
second term is called the tensor term and  operates in states with
nonzero orbital angular momentum between the quarks $i$ and $j$.

The Fermi contact term and the tensor term of order
$\alpha_s^2$~\cite{Gupta and Radford24,Gupta and
Radford26,J.Eiglsperger} derived from two-gluon exchange are
\begin{eqnarray}
H_{\mathrm{contact}}^{ij}(\alpha_s^2)&=& \frac{16\pi \alpha_s}{9
m_im_j}{\vec{S}_i}\cdot
{\vec{S}_j}\{[\frac{\alpha_s}{12\pi}(26+9\ln2)]\delta^3(\vec{r}_{ij})\nonumber\\
&&\hspace{-15pt}-\frac{\alpha_s}{24\pi^2}(33-2n_f)\overrightarrow{\nabla}^2[\frac{\ln(\mu_{\mathrm{GR}}r_{ij})+\gamma_\mathrm{E}}{r_{ij}}]\nonumber\\
&&\hspace{-15pt}+\frac{21\alpha_s}{16\pi^2}\overrightarrow{\nabla}^2[\frac{\ln(\sqrt{m_im_j}r_{ij})+\gamma_\mathrm{E}}{r_{ij}}]\},\\
H_{\mathrm{tensor}}^{ij}(\alpha_s^2)&=& \frac{2\alpha_s}{3m_im_j
}\frac{1}{r_{ij}^3}[\frac{3({\vec{S}_i}\cdot
\vec{r}_{ij})({\vec{S}_j}\cdot\vec{r}_{ij})}{r_{ij}^2}-{\vec{S}_i}\cdot{\vec{S}_j}]\cdot\nonumber\\
&&\hspace{-15pt}\{\frac{4\alpha_s}{3\pi}+\frac{\alpha_s}{6\pi}(33-2n_f)[\ln(\mu_{\mathrm{GR}}r_{ij})+\gamma_\mathrm{E}-\frac{4}{3}]\nonumber\\
&&\hspace{-15pt}
-\frac{3\alpha_s}{\pi}[\ln(\sqrt{m_im_j}r_{ij})+\gamma_\mathrm{E}-\frac{4}{3}]\},
\end{eqnarray}
where $\mu_{\mathrm{GR}}$ is a renormalization scale, the subscript
$\mathrm{GR}$ refers to the renormalization scheme~\cite{Gupta and
Radford2690} and  $n_f$ is the number of effective quark flavors.

We then write approximate solutions~\cite{Isgur and Karl2653,Isgur
and Karl1191} by perturbation theory in $U$ and $H_{\mathrm{hyp}}$.
In the $U=H_{\mathrm{hyp}}=0$ limit, $H_0$ becomes
\begin{eqnarray}
H_0\rightarrow
H_0'=\frac{p_\rho^2}{2m_\rho}+\frac{p_\lambda^2}{2m_\lambda}+\frac{3K}{2}(\rho^2+\lambda^2),
\end{eqnarray}
where
\begin{eqnarray}
\vec{\rho}&=&\frac{1}{\sqrt{2}}(\vec{r}_{1}-\vec{r_2}), \qquad
\vec{p}_{\rho}=m_\rho\frac{d\vec{\rho}}{dt},\\
\vec{\lambda}&=&\frac{1}{\sqrt{6}}(\vec{r}_{1}+\vec{r}_{2}-2\vec{r}_{3}),
\qquad \vec{p}_{\lambda}=m_\lambda\frac{d\vec{\lambda}}{dt},
\end{eqnarray}
with
\begin{eqnarray}
m_\rho&\equiv& m, \quad m_\lambda\equiv\frac{3mm'}{2m+m'},\quad
x\equiv m_d/m_s.
\end{eqnarray}
In the $S=0$ sector, $m=m'=m_d$; in the $S=-1$ sector, $m=m_d$,
$m'=m_s$; in the $S=-2$ sector, $m=m_s$, $m'=m_d$ and in the $S=-3$
sector, $m=m'=m_s$.

 The solutions to $H_0'$ are wave functions as
follows. The wave function with $N=0$ is
\begin{eqnarray}
\psi_{00}&=&\frac{\alpha_\rho^{3/2}\alpha_\lambda^{3/2}}{\pi^{3/2}}
\mathrm{exp}[-\frac{1}{2}\alpha^2_\rho\rho^2-\frac{1}{2}\alpha^2_\lambda\lambda^2)].
\end{eqnarray}
 The wave functions with $N=1$ and $N=2$ are
 \begin{eqnarray}
\psi_{11}^{\lambda}&=&\frac{\alpha_\rho^{3/2}\alpha_\lambda^{5/2}}{\pi^{3/2}}\lambda_+
\mathrm{exp}[-\frac{1}{2}\alpha^2_\rho\rho^2-\frac{1}{2}\alpha^2_\lambda\lambda^2)],\\
\psi_{11}^{\rho}&=&\frac{\alpha_\rho^{5/2}\alpha_\lambda^{3/2}}{\pi^{3/2}}\rho_+
\mathrm{exp}[-\frac{1}{2}\alpha^2_\rho\rho^2-\frac{1}{2}\alpha^2_\lambda\lambda^2)],\\
\psi_{00}^{\lambda\lambda}&=&\sqrt{\frac{2}{3}}\frac{\alpha_\rho^{3/2}\alpha_\lambda^{7/2}}{\pi^{3/2}}
(\lambda^2-\frac{3}{2}\alpha_\lambda^{-2})\mathrm{exp}[-\frac{1}{2}\alpha^2_\rho\rho^2-\frac{1}{2}\alpha^2_\lambda\lambda^2)],\\
\psi_{00}^{\rho\lambda}&=&\frac{2}{\sqrt{3}}\frac{\alpha_\rho^{5/2}\alpha_\lambda^{5/2}}{\pi^{3/2}}
\vec{\rho}\cdot\vec{\lambda}\mathrm{exp}[-\frac{1}{2}\alpha^2_\rho\rho^2-\frac{1}{2}\alpha^2_\lambda\lambda^2)],\\
\psi_{00}^{\rho\rho}&=&\sqrt{\frac{2}{3}}\frac{\alpha_\rho^{7/2}\alpha_\lambda^{3/2}}{\pi^{3/2}}
(\rho^2-\frac{3}{2}\alpha_\rho^{-2})\mathrm{exp}[-\frac{1}{2}\alpha^2_\rho\rho^2-\frac{1}{2}\alpha^2_\lambda\lambda^2)],\\
\psi_{22}^{\lambda\lambda}&=&\frac{1}{\sqrt{2}}\frac{\alpha_\rho^{3/2}\alpha_\lambda^{7/2}}{\pi^{3/2}}
\lambda_+\lambda_+\mathrm{exp}[-\frac{1}{2}\alpha^2_\rho\rho^2-\frac{1}{2}\alpha^2_\lambda\lambda^2)],\\
\psi_{22}^{\rho\lambda}&=&\frac{\alpha_\rho^{5/2}\alpha_\lambda^{5/2}}{\pi^{3/2}}
\rho_+\lambda_+\mathrm{exp}[-\frac{1}{2}\alpha^2_\rho\rho^2-\frac{1}{2}\alpha^2_\lambda\lambda^2)],\\
\psi_{22}^{\rho\rho}&=&\frac{1}{\sqrt{2}}\frac{\alpha_\rho^{7/2}\alpha_\lambda^{3/2}}{\pi^{3/2}}
\rho_+\rho_+\mathrm{exp}[-\frac{1}{2}\alpha^2_\rho\rho^2-\frac{1}{2}\alpha^2_\lambda\lambda^2)],\\
\psi_{11}^{\rho\lambda}&=&\frac{\alpha_\rho^{5/2}\alpha_\lambda^{5/2}}{\pi^{3/2}}
(\rho_+\lambda_3-\rho_3\lambda_+)\mathrm{exp}[-\frac{1}{2}\alpha^2_\rho\rho^2-\frac{1}{2}\alpha^2_\lambda\lambda^2)],
\end{eqnarray}
  where we have shown only the
highest state of an orbital angular momentum multiplet, which has
energy
$(2n_\rho+l_\rho+3/2)\omega_\rho+(2n_\lambda+l_\lambda+3/2)\omega_\lambda$
and
\begin{eqnarray}
\alpha_\rho&=&(3Km_\rho)^{1/4},\qquad\alpha_\lambda=(3Km_\lambda)^{1/4},\\
\omega_\rho&=&(3K/m_\rho)^{1/2},\qquad\omega_\lambda=(3K/m_\lambda)^{1/2}.
\end{eqnarray}

When $U$ differs from zero, it can be shown that in the $S=0$
sector, the confinement energies can be determined by three
constants $E_0$, $a$, and $b$,
\begin{eqnarray}
&&E[\psi_{00}]=E_0,\label{u1}\\
&&E[\psi_{1m}]=E_0+a,\\
&&E[\psi_{00}^{\rho\rho}]=E[\psi_{00}^{\lambda\lambda}]=E_0+2a-\frac{3}{4}b,\\
&&E[\psi_{00}^{\rho\lambda}]=E_0+2a-\frac{1}{2}b,\\
&&E[\psi_{2m}^{\rho\rho}]=E[\psi_{2m}^{\lambda\lambda}]=E_0+2a-\frac{3}{10}b,\\
&&E[\psi_{2m}^{\rho\lambda}]=E_0+2a-\frac{1}{5}b,\\
&&E[\psi_{1m}^{\rho\lambda}]=E_0+2a,\\
&&\langle\psi_{00}^{\rho\rho}|U|\psi_{00}^{\lambda\lambda}\rangle=-\frac{1}{4}b,\\
&&\langle
\psi_{2m}^{\rho\rho}|U|\psi_{2m}^{\lambda\lambda}\rangle=-\frac{1}{10}b\label{u2}.
\end{eqnarray}
To break SU(3), $\rho$ and $\lambda$ excitation energies are
decreased by $(m_d/m_\rho)^{1/2}$ and $(m_d/m_\lambda)^{1/2}$
respectively, as they would be in the harmonic limit. We use the
confinement energies shown in Table~\ref{tab:table1}, obtained by
considering $U$ effects in Refs.~\cite{Isgur and Karl1191,Kuang-Ta
Chao and Isgur 155}.

We then calculate the hyperfine matrix elements. In the nonstrange
sector, the states are completely symmetric in flavor, spin and
space under interchange of any two quarks, so it follows that
\begin{eqnarray}
\langle \alpha |H_{\mathrm{hyp}}|\beta\rangle=3\langle \alpha
|H_{\mathrm{hyp}}^{12}|\beta\rangle,
\end{eqnarray}
where $|\alpha\rangle$ and $|\beta\rangle$ are any two states. This
trick simplifies the calculation. In the $S=-1$ sector, since the
states are always symmetric under exchange of quarks one and two,
then
\begin{eqnarray}
\langle \alpha |H_{\mathrm{hyp}}^{13}|\beta\rangle=\langle \alpha
|H_{\mathrm{hyp}}^{23}|\beta\rangle,
\end{eqnarray}
and the hyperfine matrix elements become
\begin{eqnarray}
\langle \alpha |H_{\mathrm{hyp}}|\beta\rangle=\langle \alpha
|H_{\mathrm{hyp}}^{12}+2H_{\mathrm{hyp}}^{13}|\beta\rangle.
\end{eqnarray}
Calculation of the $H_{\mathrm{hyp}}^{12}(\vec{r}_{12})$ matrix
elements is straightforward. To perform the
$H_{\mathrm{hyp}}^{13}(\vec{r}_{13})$ matrix elements, we relate to
the interaction between quarks one and two by using permutations:
\begin{eqnarray}
\langle \alpha |H_{\mathrm{hyp}}^{13}|\beta\rangle=\langle \alpha
|(23)(23)H_{\mathrm{hyp}}^{13}(23)(23)|\beta\rangle,
\end{eqnarray}
with
\begin{eqnarray}\label{eq:H13}
&&(23)H_{\mathrm{hyp}}^{13}(23)=\frac{m}{m'}H_{\mathrm{hyp}}^{12},\\
&&(23)|\alpha\rangle=\sum_{\alpha'} c_{\alpha'}|\alpha'\rangle,
\end{eqnarray}
where $c_{\alpha'}$ is the coefficient of the $|\alpha'\rangle$
states determined by spin and space functions permutations. The
details of the hyperfine matrix elements calculations are discussed
in Appendix~\ref{app1}. The hyperfine matrix elements of the $S=-2$
and $S=-3$ sectors can be obtained from the $S=0$ and $S=-1$ sectors
by making the interchange $m_u\leftrightarrow m_s$ everywhere. Since
the hyperfine matrix elements of $\mathrm{O}(\alpha_s)$ have been
calculated~\cite{Isgur and Karl4187,Isgur and Karl2653,Isgur and
Karl1191}, we only calculate the hyperfine matrix elements of
$\mathrm{O}(\alpha_s^2)$  derived from two-gluon exchange. The
results are listed in Tables~\ref{table5} and \ref{table6}. The
detailed calculation of the hyperfine matrix elements of higher
order are discussed in Appendix~\ref{app2}. We also express the
hyperfine matrix elements with $\mathrm{O}(\alpha_s^2)$  here in the
unit of $\delta$, which is $4\alpha_s \alpha^3/3\sqrt{2\pi}m_d^2$.
In this paper, we assume that this value differs
slightly~\cite{Isgur and Karl2653,Isgur and Karl1191}, which is 270
MeV for considering the results of order $\alpha_s^2$ effects and
mixing between the ground states with $N=0$ and  the exited states
with $N=2$. It shows that the values of matrix elements of higher
order are in the range (0.0014-0.0334)$\delta$, i.e, the values of
the matrix elements are in the range 0.4-10 MeV, which are one order
less than those 2-170 MeV obtained by order $\alpha_s$. At last we
diagonalize the complete Hamiltonian matrices to get the baryon
masses. The hyperfine interaction of the order $\alpha_s^2$
 makes the baryon state masses mostly change by 1-25 MeV. The compositions and calculated masses are
listed in Tables~\ref{SN}, \ref{Sdelta},   \ref{S-3}, \ref{S1lamda},
\ref{S1simga}, \ref{S-2} and \ref{negative}.

\section{\label{sec:2} Results }
\subsection{The ground-state baryons}

We can find that the calculated masses of the ground states  agree
well with the experimental values~\cite{PDG}. For the ground states,
in addition, we compare our results of the order
 $\alpha_s^2$ with the results using the order
$\alpha_s$ in Table~\ref{tab:table2}. It shows that the results of
order $\alpha_s^2$  are agree better with the experimental values,
since most $\Delta M$ are somewhat smaller and the value of $\sum
(\Delta M)^2 =33$ $\mathrm{MeV}^2$ is much smaller than that of 193
$\mathrm{MeV}^2$ in Ref.~\cite{Isgur and Karl1191}.

We further study the corrections to the Gell-Mann--Okubo mass
formula (GMO) and Gell-Mann's equal spacing rule (GME). The
corrections to Gell-Mann--Okubo mass formula for the baryon-octet
and Gell-Mann's equal spacing rule for the baryon-decuplet can be
written as
\begin{eqnarray}
\frac{M_N+M_{\Xi}}{2}&=&\frac{3M_{\Lambda}+M_{\Sigma}}{4}\label{eq:GMO}+\delta_{\mathrm{GMO}},\\
M_{\Sigma^*}-M_\Delta&=&M_{\Xi^*}-M_{\Sigma^*}+\delta_{\mathrm{GME1}}\nonumber\\
&=&M_\Omega-M_{\Xi^*}+\delta_{\mathrm{GME2}},\label{eq:GM}
\end{eqnarray}
Here as $\delta_{\mathrm{GMO}}= \delta_{\mathrm{GME1}} =
\delta_{\mathrm{GME2}}= 0$, equations~(\ref{eq:GMO}) and
(\ref{eq:GM}) are the standard mass formulas which do not consider
the $H_\mathrm{hyp}$ term. However, in the real world
$\delta_{\mathrm{GMO}}=-6.75$ MeV, $\delta_{\mathrm{GME1}}=8$ MeV
and $\delta_{\mathrm{GME2}}=11$ MeV, which stand for the deviations
of the above two mass formulas from experimental data. The
deviations can be explained by a quark-mass dependent hyperfine
interaction, together with the wave function size and mixing with
excited states. In Tab.~\ref{tab:delta}, we calculate and compare
the deviations of the mass formulas. In the first row, it is the
deviations obtained from experimental data. The deviations in the
second row are results considering the one-gluon exchange process.
After considering the higher order exchange process, we list our
deviations in the last row. It shows that the correction to the
Gell-Mann--Okubo mass formula is distinctly improved and the
correction to the Gell-Mann's equal spacing rule is also getting
 slightly better. Since the calculated $\Xi^*$ mass is not very good,
there is only little improvement on the corrections to the GME.
However, the octet ground baryons are more close to reality and the
corrections to the GMO improve significantly. It implies that the
correction to the GMO is  mainly due to the hyperfine interaction of
$\mathrm{O}(\alpha_s^2)$.

\subsection{Baryon excitations}
The excited baryon masses with spin $J$ in the range 1/2-7/2 are
listed in Tables~\ref{SN}, \ref{Sdelta}, \ref{S-3}, \ref{S1lamda},
\ref{S1simga}, \ref{S-2} and \ref{negative}. We see that most of the
  masses of positive parity states  are consistent with
the experimental data. We compare the nucleon resonances
$N_{\frac{1}{2}^+}$(1440) and $N_{\frac{1}{2}^-}$(1535). In
particular the former is called the Roper resonance and it is
interpreted as the first positive parity excited state
($|N^2S'_S\frac{1}{2}^+\rangle$ or
$|N^2(56',0^+)\frac{1}{2}^+\rangle$) with the $N=2$ harmonic
oscillator band
 in the quark  models, whereas the latter is interpreted as the first negative
parity excited state ( $|N^2P_M\frac{1}{2}^-\rangle$ or
$|N^{2}(70,1^-)\frac{1}{2}^-\rangle$) with $N=1$. In the relativized
quark model with one-gluon-exchange~\cite{Capstick and Isgur} or
instanton-induced interactions ~\cite{Loring}, the calculated mass
of the $N_{\frac{1}{2}^+}$(1440) exceeds the mass of the
$N_{\frac{1}{2}^-}$(1535) by 80 MeV. It is not consistent with the
experimental value that should be 55-125 MeV below. In this work,
however, using the nonrelativistic quark model with
two-gluon-exchange, the calculated mass of $N_{\frac{1}{2}^+}$(1440)
is 1462 MeV, which locates below that of $N_{\frac{1}{2}^-}$(1535),
i.e., 1497 MeV. That can be explained as follows. First, in the
harmonic-oscillator model (when $U=0$ and $H_{\mathrm{hyp}}=0$), the
nonstrange states of SU(6) multiplets with the same harmonic
oscillator band $N$ are degenerate with $E=(N+3)\omega$. It makes
the energy of the $N=2$ positive parity excited states  1$\omega$
larger than that of the $N=1$ negative parity excited states.
However, the anharmonicity $U=\sum_{i<j} U(r_{ij})$ is a diagonal
perturbation and break the initial degeneracy within the $N=2$
harmonic oscillator band. Thus the five SU(6) multiplets at $N=2$
have different energies, which we may denote by E$(56',0^+)$,
E$(70,0^+)$, E$(56,2^+)$, E$(70,2^+)$ and E$(20,1^+)$. The energy of
$(70,1^-)$ multiplet
 at $N=1$ is expressed as E$(70,1^-)$. These energies can be expressed by three
 parameters, i.e., $a$, $b$ and $E_0$ in (\ref{u1})-(\ref{u2}).  Using the reasonable
parameters, E$(56',0^+)$ becomes 1600 MeV, which is less than
E$(70,1^-)$=1610 MeV~\cite{Isgur and Karl2653,Isgur and Karl1191}.
That is to say that the baryon energies associated with the
$(56',0^+)$ multiplet (as indicated by the Roper resonance) are less
than the baryon energies associated with the $(70,1^-)$ multiplet.
Second, the hyperfine interaction further causes splitting between
the two states. It is noted that the diagonal matrix element of the
contact term of $\mathrm{O}(\alpha_s)$ in the first positive parity
excited state $\langle N^2(56',0^+)\frac{1}{2}^+\mid
H_\mathrm{{contact}}(\mathrm{O}(\alpha_s))\mid
N^2(56',0^+)\frac{1}{2}^+\rangle$ is $-169$ MeV, while that in the
first negative parity excited state $\langle
N^2(70,1^-)\frac{1}{2}^-\mid
H_\mathrm{{contact}}(\mathrm{O}(\alpha_s))\mid
N^2(70,1^-)\frac{1}{2}^-\rangle$ is $-67$ MeV. The  splitting
between the two matrix elements is about $-100$ MeV, which further
makes the energy of the first positive parity excited state lower.
The contact term of $\mathrm{O}(\alpha_s^2)$ makes the splitting
between the two diagonal matrix elements increase by $-7$ MeV. In
addition, the hyperfine interaction also brings the off-diagonal
matrix elements, which not only cause mixing between the states with
same $N$, but also mixing between the $N=0$ and $N=2$ band states.
After diagonalizing the complete Hamiltonian matrices, we obtain the
the baryon  masses. The mass of the first positive parity state
corresponding to $N_{\frac{1}{2}^+}$(1440) is 1462 MeV. It is in the
experimental range of 1420-1470 MeV and is lower than that of the
first negative parity corresponding to $N_{\frac{1}{2}^-}$(1535).

Meanwhile, we find that the masses of the negative parity states are
similar to the results of $\mathrm{O}(\alpha_s)$~\cite{Isgur and
Karl4187}. Some predicated masses are consistent with the
experimental data. However, some controversial states, such as
$\Lambda_{\frac{1}{2}^-}$(1405), $\Lambda_{\frac{3}{2}^-}$(1520) and
$N_{\frac{1}{2}^-}$(1535), are not consistent with the experimental
data. On one hand, the hyperfine interaction of
$\mathrm{O}(\alpha_s^2)$ only makes the  masses obtained from
$\mathrm{O}(\alpha_s)$ change by 1-25 MeV. This effect is somewhat
small and difficult to make these masses agree with the experimental
data. On the other hand, these controversial states may have
interpretations by introducing new physical contents beyond the
3q-quark model~\cite{Klempt,LiuZou,Zou,Jido}.

We find that the hyperfine interaction with two-gluon exchange does
improve the calculated masses of some excited baryons, but not all
of them. Thus it is still difficult to draw a conclusion that the
quark model with the hyperfine interaction of
$\mathrm{O}(\alpha_s^2)$  describes the excited baryon spectrum
better than that of $\mathrm{O}(\alpha_s)$.

\section{\label{sec:3} Conclusions}
In this paper, we studied the baryon spectrum with hyperfine
interaction of $\mathrm{O}(\alpha_s^2)$ due to two-gluon exchange.
We find that the matrix elements of hyperfine interaction with
higher order are in the range $(0.0014-0.0334)\delta$, i.e, the
values of the matrix elements are in the range 0.4-10 MeV, which are
one order less than those matrix elements 2-170 MeV obtained by
leading order $\mathrm{O}(\alpha_s)$. After diagonalizing the
complete Hamiltonian matrices, the order $\alpha_s^2$ makes the
baryon state masses mostly change by 1-25 MeV. In addition, the
order $\mathrm{O}(\alpha_s^2)$ produces the calculated  masses of
ground states shift 7-10 MeV comparing with that of the order 60 to
200 MeV generated by $\mathrm{O}(\alpha_s)$ in term of $\delta=270$
MeV. The calculated masses are in good agreement with the
experimental masses. The value $\sum (\Delta M)^2$ shows that the
predicted masses are closer to the experimental data by taking into
account the higher order hyperfine interaction. In addition, the
correction to the Gell-Mann--Okubo mass formula for the baryon-octet
is distinctly improved and the correction to the Gell-Mann's equal
spacing rule for the baryon-decuplet is also slightly improved.

It is noted that $\alpha_s$ becomes 0.7 when considering higher
order hyperfine interactions, and it is smaller than the value
$\pi$/2 in Ref.~\cite{Isgur and Karl1191}. This implies that the
applicability of perturbative QCD calculations becomes more
reasonable in the quark-model framework by taking into account
higher order effects. In addition, we employ the same value of the
parameter $\delta$ for both ground and excited baryons. We assume
that the value is about 270 MeV to consider the effects of the
hyperfine interactions with higher order $\alpha_s^2$ and the mixing
between ground states and excited states. It is different from that
the $\delta$ value of 300 MeV for the excited states in
Ref.~\cite{Isgur and Karl2653} and  260 MeV for the ground states in
Ref.~\cite{Isgur and Karl1191}. Some parameters used in this paper
are the same as those used in Ref.~\cite{Isgur and Karl2653,Isgur
and Karl1191,Kuang-Ta Chao and Isgur 155}. Other parameters are
roughly derived in terms of dynamical theory. These parameters can
change in some reasonable ranges, but cannot change the order of
magnitude of the physical quantities.

 Finally we conclude that the effects from higher order hyperfine interaction
  should be considered in the quark model. Taking into account
the hyperfine interaction of $\mathrm{O}(\alpha_s^2)$ due to
two-gluon exchange, the calculated spectrum agree better with the
experimental data. The higher order hyperfine interaction bring also
more realistic correction to the Gell-Mann--Okubo mass formula.
Corrections due to two-gluon exchange on baryon excitations
(including negative parity baryons) are also briefly discussed.

\bigskip
\vspace{0.5cm}

We are grateful to Bin Wu, Fanyong Zou and Wen Qian for useful
discussions. This work is partially supported by National Natural
Science Foundation of China (No.~10721063), by the Key Grant Project
of Chinese Ministry of Education (No.~305001), and by the Research
Fund for the Doctoral Program of Higher Education (China).
\appendix
\renewcommand{\theequation}{A\arabic{equation}}
\setcounter{equation}{0}
\section{\label{app1}The hyperfine matrix elements in the SU(6) basis}

 In this appendix we
present detailed calculation of the hyperfine matrix elements of
$\Lambda$ in the SU(6) and $uds$ bases.

The spectrum of $N$, $\Delta$ and $\Omega$ can be easily calculated
in the SU(6) basis. However,  $\Sigma$,  $\Lambda$ and $\Xi$  are
usually first calculated by the matrix in the $uds$ basis, which are
simpler. Then using the relation between the  SU(6) and $uds$ bases,
we transform the matrix in the $uds$ basis to the matrix in the
SU(6) basis and obtain the baryon spectrum and the composition of
the eigenstates in terms of the SU(6) basis.

Take $\Lambda$ for example,  the SU(6) wave functions are \\
(octet)
\begin{eqnarray}
\Phi_\Lambda^\rho&=&-\frac{1}{\sqrt{12}}(2uds-2dus+usd-dsu-sud+sdu),\nonumber\\\\
\Phi^\lambda_\Lambda&=&-\frac{1}{2}(usd-dsu+sud-sdu),
\end{eqnarray}
(singlet)
\begin{eqnarray}
\Phi_\Lambda^A&=&\frac{1}{\sqrt{6}}(uds-dus-usd+dsu+sud-sdu).\nonumber\\
\end{eqnarray}
The $uds$ type wave function is
\begin{eqnarray}
\Phi_\Lambda=\frac{1}{\sqrt{2}}(ud-du)s,
\end{eqnarray}
and the spin wave functions are
\begin{eqnarray}
\chi_+^\rho&=&\frac{1}{\sqrt{2}}(\uparrow \downarrow \uparrow-\downarrow\uparrow\uparrow),\\
\chi_-^\rho&=&\frac{1}{\sqrt{2}}(\uparrow \downarrow \downarrow-\downarrow\uparrow\downarrow),\\
\chi_+^\lambda&=&-\frac{1}{\sqrt{6}}(\uparrow\downarrow\uparrow+\downarrow\uparrow\uparrow-2\uparrow\uparrow\downarrow),\\
\chi_-^\lambda&=&\frac{1}{\sqrt{6}}(\uparrow\downarrow\downarrow+\downarrow\uparrow\downarrow-2\downarrow\downarrow\uparrow),\\
\chi_{3/2}^S&=&\uparrow\uparrow\uparrow, ~~\mathrm{ etc}.
\end{eqnarray}
The $uds$-type $\Lambda$ states
\begin{eqnarray}
|\Lambda^2S_{\lambda\lambda}\frac{1}{2}^+\rangle&=&\phi_\Lambda\chi_+^\rho\psi_{00}^{\lambda\lambda},\\
|\Lambda^2S_{\rho\rho}\frac{1}{2}^+\rangle&=&\phi_\Lambda\chi_+^\rho\psi_{00}^{\rho\rho},\\
|\Lambda^2S_{\rho\lambda}\frac{1}{2}^+\rangle&=&\phi_\Lambda\chi_+^\lambda\psi_{00}^{\rho\lambda}.
\end{eqnarray}
We display here only the top state of a given $J^p$ multiplet, and
we also display only the state of maximum $J$ from coupling a given
$L$ and $S$; the states of smaller $J$ are constructed using
standard tables in the $LS$ order.

The relation between the SU(6) and $uds$ bases is that
\begin{eqnarray}
\left(\begin{array}{c}\Lambda^2L_{\lambda\lambda}\\
\Lambda^2L_{\rho\rho}\\
\Lambda^2L_{\rho\lambda}\end{array}\right)\leftrightarrow
\left(\begin{array}{rrr}
\frac{1}{\sqrt{2}} & \frac{1}{2} &-\frac{1}{2}\\
\frac{1}{\sqrt{2}}& -\frac{1}{2}  & \frac{1}{2}\\
0&\frac{1}{\sqrt{2}} & \frac{1}{\sqrt{2}}
\end{array}\right)
\left(\begin{array}{c}
{\Lambda_8}^2L_S\\
{\Lambda_1}^2L_M\\
{\Lambda_8}^2L_M\end{array}\right),
\end{eqnarray}
where we use $\Lambda$ instead of  $|\Lambda\rangle$ to denote the
state for simplicity.
\begin{eqnarray}
|\Lambda^4L_{\rho\lambda}\rangle\leftrightarrow|{\Lambda_8}^4L_M\rangle,
\end{eqnarray}
for $L=S$ or $D$ and
\begin{eqnarray}
|\Lambda^2P_{\rho\lambda}\rangle\leftrightarrow-|{\Lambda_8}^2P_A\rangle,\\
|\Lambda^4P_{\rho\lambda}\rangle\leftrightarrow-|{\Lambda_1}^4P_A\rangle.
\end{eqnarray}
so the $H_{\mathrm{hyp}}$ matrix elements of the SU(6) and $uds$
bases have the relation
\begin{eqnarray}
\langle\alpha_{SU(6)}|H_{\mathrm{hyp}}^{\mathrm{SU(6)}}|\beta_{SU(6)}\rangle=\langle\alpha_{uds}|H_{\mathrm{hyp}}^{\mathrm{uds}}|\beta_{uds}\rangle,
\end{eqnarray}
where
\begin{eqnarray}
\beta_{uds}&=&S^+\beta_{SU(6)},\\
H_{\mathrm{hyp}}^{\mathrm{SU(6)}}&=&SH_{\mathrm{hyp}}^{\mathrm{uds}}S^+.
\end{eqnarray}
Take the $\Lambda$ with  $J^p=1/2^+$ with $N=2$ for example, the
relation between the SU(6) and $uds$ bases is
\begin{eqnarray}
\left(\begin{array}{c}\Lambda^2S_{\lambda\lambda}\frac{1}{2}^+\\
\Lambda^2S_{\rho\rho}\frac{1}{2}^+\\
\Lambda^2S_{\rho\lambda}\frac{1}{2}^+\\
\Lambda^4D_{\rho\lambda}\frac{1}{2}^+\\
\Lambda^2P_{\rho\lambda}\frac{1}{2}^+\\
\Lambda^4P_{\rho\lambda}\frac{1}{2}^+
\end{array}\right)
=
S^+ \left(\begin{array}{l}{\Lambda_8}^2S'_S\frac{1}{2}^+\\
 {\Lambda_8}^2S_M\frac{1}{2}^+\\{\Lambda_1}^2S_M\frac{1}{2}^+\\{\Lambda_8}^4D_M \frac{1}{2}^+\\
 {\Lambda_1}^4P_A \frac{1}{2}^+\\{\Lambda_8}^2P_A \frac{1}{2}^+
 \end{array}\right),
\end{eqnarray}
where
\begin{eqnarray}\label{eq:relation}
S^+=\left(\begin{array}{cccccc}
+\frac{1}{\sqrt{2}} & -\frac{1}{2} &\frac{1}{2} &0&0&0\\
\frac{1}{\sqrt{2}}& \frac{1}{2}  &-\frac{1}{2} &0&0&0\\
0&+\frac{1}{\sqrt{2}} & +\frac{1}{\sqrt{2}} &0&0&0\\
0                  &0                     &0                               &+1&0&0\\
0                  &0                     &0 &0&-1&0\\
0                  &0                     &0 &0&0&-1\\
\end{array}\right).
\end{eqnarray}
The matrix $H_\mathrm{contact}$ in the $uds$ basis with order
$\alpha_s$ is
\begin{eqnarray}
\hspace{-15pt}H_\mathrm{contact}=\delta\left(\begin{array}{cccccc}
-\frac{1}{2}            & 0                    &-\frac{7\sqrt{2}}{32}x            &0&0&0\\
0                       & -\frac{3}{4}          &+\frac{3\sqrt{2}}{32}x           &0&0&0\\
-\frac{7\sqrt{2}}{32}x  &+\frac{3\sqrt{2}}{32}x &- \frac{5}{8}x                  &0&0&0\\
0                  &0                     &0                               &\frac{1}{8}x &0&0\\
0&0&0 &0&0&0\\
0&0&0 &0&0&0\\
\end{array}\right).
\end{eqnarray}
Using the relation in Eq.~(\ref{eq:relation}), the matrix
$H'_\mathrm{contact}$ in the SU(6) bases is
\begin{eqnarray}
&&H_{\mathrm{contact}}'=SH_{\mathrm{contact}}S^+=\\
&& \hspace{-30pt}\delta\left(\begin{array}{cccccc}
-\frac{5}{8}                  &-\frac{\sqrt{2}}{16}(1+x)           &+\frac{\sqrt{2}}{16}(1-x)            &0&0&0\\
-\frac{\sqrt{2}}{16}(1+x)     & -\frac{5}{16}                      &+\frac{5}{16}(1-x)           &0&0&0\\
+\frac{\sqrt{2}}{16}(1-x)     &+\frac{5}{16}(1-x)                 &-\frac{5}{16}(1+2x)                &0&0&0\\
0                  &0                     &0                               &\frac{1}{8}x &0&0\\
0&0&0 &0&0&0\\
0&0&0 &0&0&0
\end{array}\right).\nonumber
\end{eqnarray}
In addition, we show some detailed calculation of the $H^{13}$
matrix elements. As mentioned in Eq.~(\ref{eq:H13}), under exchange
of quarks two and three, i.e., the relative coordinates become
\begin{eqnarray}
(23)\rho=\frac{1}{2}\rho+\frac{\sqrt{3}}{2}\lambda,\\
(23)\lambda=\frac{\sqrt{3}}{2}\rho-\frac{1}{2}\lambda,
\end{eqnarray}
the spin functions are
\begin{eqnarray}
&&(23)\chi^S=\chi^S,\\
&&(23)\chi^\rho=\frac{1}{2}\chi^\rho+\frac{\sqrt{3}}{2}\chi^\lambda,\\
&&(23)\chi^\lambda=\frac{\sqrt{3}}{2}\chi^\rho-\frac{1}{2}\chi^\lambda.
\end{eqnarray}
In term of the approximation $\alpha_\rho\simeq\alpha_\lambda$, the
space functions are
\begin{eqnarray}
(23)\psi_{00}&=&\psi_{00},\\
(23)\psi_{00}^{\lambda\lambda}&=&\frac{3}{4}\psi_{00}^{\rho\rho}+\frac{1}{4}\psi_{00}^{\lambda\lambda}-\frac{\sqrt{6}}{4}\psi_{00}^{\rho\lambda},\\
(23)\psi_{00}^{\rho\rho}&=&\frac{1}{4}\psi_{00}^{\rho\rho}+\frac{3}{4}\psi_{00}^{\lambda\lambda}+\frac{\sqrt{6}}{4}\psi_{00}^{\rho\lambda},\\
(23)\psi_{00}^{\rho\lambda}&=&\frac{\sqrt{6}}{4}\psi_{00}^{\rho\rho}-\frac{\sqrt{6}}{4}\psi_{00}^{\lambda\lambda}+\frac{1}{2}\psi_{00}^{\rho\lambda},\\
(23)\psi_{11}^{\rho\lambda}&=& -\psi_{11}^{\rho\lambda},
\end{eqnarray}
where $(23)\psi_{22}^{\lambda\lambda}$, $(23)\psi_{22}^{\rho\rho}$
and  $(23)\psi_{22}^{\rho\lambda}$ are similar  with those of
$(23)\psi_{00}^{\lambda\lambda}$, $(23)\psi_{00}^{\rho\rho}$ and
$(23)\psi_{00}^{\rho\lambda}$.

 Therefore, if $|\alpha\rangle=|\beta\rangle=|\Lambda^2
 S_{\rho\lambda}1/2^+\rangle$, the matrix element $H^{13}_{\mathrm{contact}}$ is
\begin{eqnarray}
&&\langle\chi_+^\lambda\psi_{00}^{\rho\lambda}|H_{\mathrm{contact}}^{13}|\chi_+^\lambda\psi_{00}^{\rho\lambda}\rangle
=
\langle(23)\psi_{00}^{\rho\lambda}|x\delta^3(\vec{r}_{12})|(23)\psi_{00}^{\rho\lambda}\rangle\nonumber\\
&& \hspace{80pt}\times\frac{16\pi
\alpha_s}{9m_d^2}\langle(23)\chi_+^\lambda|\vec{S}_{1}\cdot
\vec{S}_{2}|(23)\chi_+^\lambda\rangle\nonumber\\
&=& x \frac{16\pi \alpha_s}{9m_d^2}
[\frac{6}{16}\langle\psi_{00}^{\rho\rho}|\delta^3(\vec{r}_{12})|\psi_{00}^{\rho\rho}\rangle \nonumber\\
&&+\frac{6}{16}\langle\psi_{00}^{\lambda\lambda}|\delta^3(\vec{r}_{12})|\psi_{00}^{\lambda\lambda}\rangle]\cdot\frac{-1}{2}\nonumber\\
 &=&\frac{-5\alpha_s \alpha^3}{12\sqrt{2\pi}m_d^2}x=\frac{-5x}{16}\delta.
\end{eqnarray}

\renewcommand{\theequation}{B\arabic{equation}}
\setcounter{equation}{0}
\section{\label{app2} The hyperfine matrix elements with order $\alpha_s^2$ }

The hyperfine matrix elements with order  $\alpha_s$ have been
calculated~\cite{Isgur and Karl4187,Isgur and Karl2653}. We
calculate the matrix elements with order $\alpha_s^2$  by the same
method. The simplest case corresponds to the interaction between
quarks 1 and 2~\cite{Capstick and Isgur,J.Eiglsperger}.
 The contact matrix elements are:
\begin{eqnarray}
&&\langle\alpha|
H_{\mathrm{contact}}^{12}|\beta\rangle=\delta_{L_\alpha
L_\beta}\delta_{l_{\rho_\alpha }l_{\rho_\beta
}}\delta_{n_{\lambda_\alpha }n_{\lambda_\beta
}}\delta_{l_{\lambda_\alpha }l_{\lambda_\beta }}\nonumber\\
&&\times\langle S_\alpha S_m|\vec{S}_{1}\cdot\vec{S}_{2}|S_\beta
S_m\rangle \frac{16\pi \alpha_s}{9 m^2}\nonumber\\
&& \langle
n_{\rho_\alpha}l_{\rho_\alpha}|[\frac{\alpha_s}{12\pi}(26+9\ln2)]\delta^3(\sqrt{2}\vec{\rho})
\nonumber\\&&-\frac{\alpha_s}{24\pi^2}(33-2n_f)\overrightarrow{\nabla}^2[\frac{\ln(\mu_{GR}\sqrt{2}\rho)+\gamma_E}{r_{ij}}]
\nonumber\\&&+\frac{21\alpha_s}{16\pi^2}\overrightarrow{\nabla}^2[\frac{\ln(m\sqrt{2}\rho)+\gamma_E}{\sqrt{2}\rho}]|
n_{\rho_\beta}l_{\rho_\beta}\rangle. \nonumber\\
\end{eqnarray}
The tensor term:
\begin{eqnarray}
H_{\mathrm{tensor}}^{12}=V(\sqrt{2}\rho)\mathbf{R}_{2}(12)\cdot
\mathbf{S}_{2}(12),
\end{eqnarray}
where $\mathbf{R}_{2}(12)$ and $\mathbf{S}_{2}(12)$ are the
spatial-tensor and  spin-tensor operators respectively.

The calculation of the tensor matrix elements are done with the aid
of the Wigner-Eckart theorem~\cite{Brinkand,Varshalovich}, which is
applied twice: once to the scalar product of the spin-tensor and
spatial-tensor operators, and the other to the $L=2$ tensor operator
where $L$ is made up from coupling $l_\rho$ and $l_\lambda$.

 We apply the Wigner-Eckart theorem to the tensor product $H_{\mathrm{tensor}}^{12}$ and obtain
\begin{eqnarray}\label{eq:once wigner}
&&\langle\alpha|
H_{\mathrm{tensor}}^{12}|\beta\rangle=(-1)^{J-L_\alpha-S_\beta}W(L_\alpha
L_\beta S_\alpha S_\beta;2J
)\nonumber\\&&\times\sqrt{2L_\alpha+1}\sqrt{2S_\alpha+1}\langle
S_\alpha ||\mathbf{S}_{2}(12)||S_\beta \rangle\\
&&\times \langle L_\alpha n_{\rho_\alpha}l_{\rho_\alpha}
n_{\lambda_\alpha }
l_{\rho_\alpha}||V(\sqrt{2}\rho)\mathbf{R}_{2}(12)||L_\beta
n_{\rho_\beta}l_{\rho_\beta} n_{\lambda_\beta} l_{\rho_\beta}\rangle
.\nonumber
\end{eqnarray}
The theorem is applied again to the spatial reduced matrix element
\begin{eqnarray}\label{eq:twice wigner}
&&\langle L_\alpha n_{\rho_\alpha}l_{\rho_\alpha} n_{\lambda_\alpha
} l_{\rho_\alpha}||V(\sqrt{2}\rho)\mathbf{R}_{2}(12)||L_\beta
n_{\rho_\beta}l_{\rho_\beta} n_{\lambda_\beta}
l_{\rho_\beta}\rangle\nonumber\\&&=(-1)^{L_\alpha+l_{\rho_{\beta}}-2-l_{\lambda_{\alpha}}}\delta_{n_{\lambda_\alpha
}n_{\lambda_\beta }}\delta_{l_{\lambda_\alpha }l_{\lambda_\beta }}
\nonumber\\&&\times W(l_{\rho_{\alpha}}l_{\rho_\beta}L_\alpha
L_\beta;2l_{\lambda_\alpha})\sqrt{2L_\beta+1}\sqrt{2l_{\rho_\alpha}+1}
\nonumber\\&&\times
 \langle n_{\rho_\alpha}l_{\rho_\alpha}||V(\sqrt{2}\rho)\mathbf{R}_{2}(12)||
n_{\rho_\beta}l_{\rho_\beta} \rangle,
\end{eqnarray}
where
\begin{eqnarray}
 &&\langle
n_{\rho_\alpha}l_{\rho_\alpha}||V(\sqrt{2}\rho)\mathbf{R}_{2}(12)||
n_{\rho_\beta}l_{\rho_\beta}
\rangle=\\&&\hspace{-10pt}C(l_{\rho_\beta}200;l_{\rho_\alpha}0)\left[\frac{2(2l_{\rho_\beta}+1)}{2l_{\rho_\alpha}+1}\right]^{1/2}\langle
n_{\rho_\alpha}l_{\rho_\alpha}|V(\sqrt{2}\rho)|
n_{\rho_\beta}l_{\rho_\beta} \rangle,\nonumber
\end{eqnarray}
and the hyperfine tensor energy $V(\sqrt{2}\rho)$ with order
$\alpha_s^2$ is
\begin{eqnarray}
&&\hspace{-25pt}V(\sqrt{2}\rho)=\frac{2\alpha_s}{3m^2
}\frac{1}{(\sqrt{2}\rho)^3}\{\frac{4\alpha_s}{3\pi}+\frac{\alpha_s}{6\pi}(33-2n_f)\\
&&\hspace{-25pt}\times[\ln(\mu_{GR}\sqrt{2}\rho)+\gamma_E-\frac{4}{3}]
-\frac{3\alpha_s}{\pi}[\ln(m\sqrt{2}\rho)+\gamma_E-\frac{4}{3}]\}.\nonumber
\label{eq:b2}
\end{eqnarray}

\bigskip
\vspace{0.5cm}

 For comparing with the hyperfine matrix elements with
order $\alpha_s$, we also express the hyperfine matrix elements with
order $\alpha_s^2$ in the unit of $\delta$, which is
\begin{eqnarray}\label{eq:eqB}
\delta \equiv\frac{4\alpha_s \alpha^3}{3\sqrt{2\pi}m_d^2}\simeq 270~
\mathrm{MeV} .
\end{eqnarray}
The $\delta$ value of 300~MeV in Ref.~\cite{Isgur and Karl2653} is
determined by considering only exited states with $N=2$. When
considering the effects of mixing between  ground states with $N=0$
and exited states with $N=2$ in Ref.~\cite{Isgur and Karl1191}, the
$\delta$ value is assumed to be 260 MeV. In this paper, we assume
that this value differs slightly, which is $\delta\simeq 270~
\mathrm{MeV}$, to consider the results of order $\alpha_s^2$ effects
and mixing between the ground states with $N=0$ and the excited
states with $N=2$.

In the calculation of the hyperfine matrix elements with order
$\alpha_s^2$, there are more parameters needed. They are $\alpha_s$,
$\mu_{GR}$, $m_d$, $\alpha$, $n_f$ and $x$, which satisfy
$\delta\simeq 270 ~\mathrm{MeV}$. Since it is hard to derive the
precise values of the parameters from foundational principle, we
assume the reasonable values in term of the known ranges of the
parameters. The light baryons belong to the soft regime. It has been
shown~\cite{A.C.Aguilar}, through a solution of Schwinger-Dyson
equation, that when decreasing $Q^2$, $\alpha_s$ does not increase
continuously but instead it saturates at a critical value
$\alpha_s(0)\approx0.7\pm0.3$. In our study regime, we assume
$\alpha_s=0.7$. The effective mass range of the constituent quark is
$m_d\simeq200-350$~MeV, and we suppose $m_d=0.275~\mathrm{GeV}$ here
and the parameter $\alpha\approx0.38$ to satisfy the value of
Eq.~(\ref{eq:eqB}).  By the similar relation that
$\mu_{c\mathrm{GR}}\propto m_c$ for charmonium in
Ref.~\cite{J.Eiglsperger}, we suppose that the renormalization scale
$\mu_{d\mathrm{GR}}=0.333 ~\mathrm{GeV}$ for baryons with
$m_1=m_2=m_d$, and $\mu_{s\mathrm{GR}}=0.333/x~\mathrm{GeV}$ for
baryons with $m_1=m_2=m_s$, where we  take  $x=0.616$.  The number
of effective quark flavors is $n_f=2$, which is the number of quarks
with mass less than the energy scale $\mu_{\mathrm{GR}}$.

In summary, we use the values $\alpha_s=0.7$, $x=0.616$,
$\alpha=0.38$,  $m_d=0.275\ \mathrm{GeV}$,
$\mu_{d\mathrm{GR}}=0.333\ \mathrm{GeV}$,
$\mu_{s\mathrm{GR}}=0.333/x\ \mathrm{GeV}$ and $\delta=270\
\mathrm{MeV}$. The hyperfine matrix elements are all listed in
Tables~\ref{table5} and \ref{table6}.


\newpage
\begin{table}[h]\caption{\label{tab:table1}The confinement energy in MeV.}
\begin{center}
\begin{tabular}{lcccc}
\hline\hline
& & Confinement &energy&\\
&$S=0$& $S=-1$& $S=-2$& $S=-3$\\
\hline
 $E[\psi_{00}^0]$                 &1135    &1296 &1457 &1618\\
 $E[\psi_{1m}^\rho]$              &1610   &1770 &1825 &1985\\
 $E[\psi_{1m}^\lambda]$           &1610   &1700 &1895 &1985\\
 $E[\psi_{00}^{\rho\rho}]$        &1705    &1895 &1910 &2070\\
 $E[\psi_{00}^{\rho\lambda}]$     &1810    &1945 &2040 &2150\\
 $E[\psi_{00}^{\lambda\lambda}]$  &1705    &1805 &2000 &2070\\
 $\langle\psi_{00}^{\rho\rho}|H_0|\psi_{00}^{\lambda\lambda}\rangle$
 &-105&-100&-90&-85\\
 $E[\psi_{1m}^{\rho\lambda}]$     &2020    &2145 &2225 &2315\\
 $E[\psi_{2m}^{\rho\rho}]$        &1890    &2085 &2055 &2215\\
 $E[\psi_{2m}^{\rho\lambda}]$        &1935    &2065 &2150 &2245\\
 $E[\psi_{2m}^{\lambda\lambda}]$        &1890    &1975 &2175 &2215\\
 $\langle\psi_{2m}^{\rho\rho}|H_0|\psi_{2m}^{\lambda\lambda}\rangle$
 &-40&-40&-35&-35\\
 \hline\hline
\end{tabular}
\end{center}
\end{table}

\newpage
\begin{table}[h]\caption{\label{tab:table2}The ground-state baryons (in units of MeV).}
\begin{center}
\begin{tabular}{cccccc}
\hline\hline
State ($J^P$)& Mass &Mass~\cite{Isgur and Karl1191}& $M_{\mathrm{exp}}$ & $\Delta M$&$\Delta M$~\cite{Isgur and Karl1191}\\
\hline
 $N\frac{1}{2}^+$        &939   &940  &939& 0&-1\\
 $\Lambda\frac{1}{2}^+$  &1113  &1110 &1116&3&6\\
 $\Sigma\frac{1}{2}^+$   &1192  &1190 &1193&1&3\\
 $\Xi\frac{1}{2}^+$      &1317  &1325 &1318&-1&-7\\
 $\Delta\frac{3}{2}^+$   &1231  &1240 &1232&1&-8\\
 $\Sigma^*\frac{3}{2}^+$   &1383  &1390 &1385&2&-5\\
 $\Xi^*\frac{3}{2}^+ $     &1526  &1530 &1530&4&0\\
 $\Omega\frac{3}{2}^+$    &1673  &1675 &1672&-1&-3\\
 \hline\hline
\end{tabular}
\\where the value $\Delta M=\mathrm{Mass}-M_{\mathrm{exp}}$.
\end{center}
\end{table}

\newpage
\begin{table}[h]\caption{\label{tab:delta}The corrections to GMO and GME in MeV.}
\begin{center}\begin{tabular}{cccc}
\hline\hline
& \hspace{20pt}$\delta_{\mathrm{GMO}}$  & \hspace{20pt}$\delta_{\mathrm{GME1}}$&\hspace{20pt}$\delta_{\mathrm{GME2}}$  \\
\hline
$\delta_{\mathrm{exp}}$     &\hspace{20pt}-6.75   &\hspace{20pt}8  &\hspace{20pt}11\\
$\delta$\cite{Isgur and Karl1191} &\hspace{20pt}2.5   &\hspace{20pt}10 &\hspace{20pt}5 \\
$\delta_{\mathrm{our}}   $&\hspace{20pt}-4.75  &\hspace{20pt}9 &\hspace{20pt}5\\
\hline\hline
\end{tabular}
\end{center}\end{table}

\newsavebox{\N}
\begin{lrbox}{\N}
\begin{tabular}{cccll}
\hline\hline
State&Mass&Exp.&       \hspace{20mm}Composition                           &\\
   & (MeV)         & (MeV)    & &\\
 \hline
$\begin{array}{c}N\frac{1}{2}^+\\N\frac{1}{2}^+\\N\frac{1}{2}^+\\N\frac{1}{2}^+\\N\frac{1}{2}^+\\
 \end{array}$&$\begin{array}{c}939\\1462\\1748\\1887\\2060
 \end{array}$&$\begin{array}{c}939\\1420-1470\\1680-1740\\1885-2125\\\\
 \end{array}$
 &$\left[\begin{array}{rrrrr}
  0.94 & 0.25& 0.22&-0.01& 0.00\\
  -0.26& 0.96& 0.04& 0.02& 0.00\\
  0.19 & 0.08&-0.90& 0.37 & 0.09\\
 -0.06 & -0.05& 0.37& 0.83&0.40\\
  0.01 & 0.01& -0.07&-0.41&0.91\\
 \end{array}\right]
 $&$\begin{array}{l}N^2S_S\frac{1}{2}^+\\N^2S'_S\frac{1}{2}^+\\N^2S_M\frac{1}{2}^+\\N^4D_M\frac{1}{2}^+\\N^2P_A \frac{1}{2}^+\\
 \end{array}$\\

$\begin{array}{c}N\frac{3}{2}^+\\N\frac{3}{2}^+\\N\frac{3}{2}^+\\N\frac{3}{2}^+\\N\frac{3}{2}^+\\
 \end{array}$&$\begin{array}{c}1734\\1857\\1952\\1975\\2058
 \end{array}$&$\begin{array}{c}1700-1750\\1855-1975\\\\\\\\
 \end{array}$
 &$\left[\begin{array}{rrrrr}
 -0.16& 0.86& -0.49& 0.01&0.00\\
-0.78& -0.30& -0.25& 0.46& -0.13\\
 0.57& -0.07& -0.29& 0.64& -0.41\\
 0.16& -0.41& -0.77& -0.03& 0.31\\
 0.10& 0.07& 0.10& 0.51   & 0.84\\
 \end{array}\right]
 $&$\begin{array}{l}N^4S_M\frac{3}{2}^+\\N^2D_S\frac{3}{2}^+\\N^2D_M\frac{3}{2}^+\\N^4D_M \frac{3}{2}^+\\N^2P_A\frac{3}{2}^+
 \end{array}$\\
 $\begin{array}{c}N\frac{5}{2}^+\\N\frac{5}{2}^+\\N\frac{5}{2}^+
 \end{array}$&$\begin{array}{c}1738\\1953\\2007
 \end{array}$&$\begin{array}{c}1680-1690\\\\\\
 \end{array}$
 &$\left[\begin{array}{rrr}
 0.88& -0.47& 0.00\\
 -0.45& -0.84& -0.31\\
 -0.15& -0.27& 0.95\\
 \end{array}\right]
 $&$\begin{array}{l}N^2D_S\frac{5}{2}^+\\N^2D_M\frac{5}{2}^+\\N^4D_M\frac{5}{2}^+
 \end{array}$\\

$N\frac{7}{2}^+$ &1943& $1940-2090$& 1.00&$N^4D_M\frac{7}{2}^+$\\
 \\
$\begin{array}{c} N\frac{1}{2}^-\\N\frac{1}{2}^-
 \end{array}$&$\begin{array}{c}1497\\1650
 \end{array}$&  $\begin{array}{c}1525-1545\\1645-1670
 \end{array}$   &$\left[\begin{array}{rr}
 0.80& 0.59\\
 -0.59& 0.80\\
 \end{array}\right] $&$\begin{array}{c}N^{2}P_M\frac{1}{2}^-\\N^{4}P_M\frac{1}{2}^-
 \end{array}$\\

 $\begin{array}{c}N\frac{3}{2}^-\\N\frac{3}{2}^-
 \end{array}$&$\begin{array}{c}1548\\1731
 \end{array}$&$\begin{array}{c}1515-1525\\1650-1750
 \end{array}$&$\left[\begin{array}{cc}
 0.99&-0.13\\
 0.13& 0.99\\
 \end{array}\right]
 $&$\begin{array}{c}N^{2}P_M\frac{3}{2}^-\\N^{4}P_M\frac{3}{2}^-
 \end{array}$\\
 $N\frac{5}{2}^-$ &1655&$\begin{array}{c}1670-1680
 \end{array}$ &~1.00&$~N^4P_M\frac{5}{2}^-$\\
\hline\hline
\end{tabular}
\end{lrbox}

\newpage
\begin{table}[b]
\caption{The calculated N spectrum  and composition in the SU(6)
basis. }\label{SN} \scalebox{0.8}{\usebox{\N}}
\end{table}

\newsavebox{\tdelta}
\begin{lrbox}{\tdelta}
\begin{tabular}{cccll}
\hline\hline
 State&Mass&Exp.&       \hspace{20mm}Composition                           &\\
   & (MeV)         & (MeV)    &  &\\
 \hline
$\begin{array}{c}\Delta\frac{1}{2}^+\\\Delta\frac{1}{2}^+
 \end{array}$&$\begin{array}{c}1867\\1900
 \end{array}$&$\begin{array}{c}\\1870-1920\\
 \end{array}$
 &$\left[\begin{array}{rr}
 0.56& 0.83\\
 0.83& -0.56\\
 \end{array}\right] $&$\begin{array}{r}\Delta^{2}S_M\frac{1}{2}^+\\\Delta^{4}D_S\frac{1}{2}^+
 \end{array}$\\

$\begin{array}{c}\Delta\frac{3}{2}^+\\\Delta\frac{3}{2}^+\\\Delta\frac{3}{2}^+\\\Delta\frac{3}{2}^+\\
\end{array}$&$\begin{array}{c}1231\\ 1790\\  1904\\  1972
 \end{array}$&$\begin{array}{c}1232\\1550-1700\\ 1900-1970\\ \\
 \end{array}$
 &$\left[\begin{array}{rrrr}
 0.98&- 0.21& 0.00&0.00\\
 0.21& 0.96& 0.13&-0.08\\
 0.03& 0.15& -0.92&0.37\\
 0.00& 0.02& 0.37&0.93\\
\end{array}\right]
$&$\begin{array}{l}\Delta^{4}S_S\frac{3}{2}^+\\\Delta
^{4}S'_S\frac{3}{2}^+\\\Delta^{4}D_S\frac{3}{2}^+\\\Delta^{2}D_M\frac{3}{2}^+\end{array}$\\

$\begin{array}{c}\Delta\frac{5}{2}^+\\\Delta\frac{5}{2}^+
 \end{array}$&$\begin{array}{c}1931\\1967
 \end{array}$&$\begin{array}{c}1865-1915\\1724-2200
 \end{array}$
 &$\left[\begin{array}{cc}
 0.92& 0.39\\
 - 0.39&0.92\\
 \end{array}\right] $&$\begin{array}{l} \Delta^4D_S\frac{5}{2}^+\\\Delta^2D_M\frac{5}{2}^+
 \end{array}$\\

$\Delta\frac{7}{2}^+$ &1902&$1915-1950$
 &1.00&$\Delta^4D_S\frac{7}{2}^+$\\
 \\

$\Delta\frac{1}{2}^-$ &1668&$1600-1660$&~1.00&$~\Delta^2P_M\frac{1}{2}^-$\\

$\Delta\frac{3}{2}^-$ &1668&$1670-1750$&~1.00&$~\Delta^2P_M\frac{3}{2}^-$\\
\hline\hline
\end{tabular}
\end{lrbox}

\newpage
\begin{table}[h]
\caption{The calculated  $\Delta$ spectrum  and composition in the
SU(6) basis. }\label{Sdelta}
\scalebox{1}{\usebox{\tdelta}}\end{table}

\newsavebox{\tomege}

\begin{lrbox}{\tomege}
\begin{tabular}{cccll}
 \hline\hline
  State&Mass&Exp.&       \hspace{20mm}Composition                           &\\
    & (MeV)         & (MeV)    &  &\\
 \hline
$\begin{array}{c}\Omega\frac{1}{2}^+\\\Omega\frac{1}{2}^+
 \end{array}$&$\begin{array}{c}2182\\2202
 \end{array}$&$\begin{array}{c}\\\\
 \end{array}$
 &$\left[\begin{array}{rr}
 0.61& 0.79\\
 0.79& -0.61\\
 \end{array}\right] $&$\begin{array}{r}\Omega^{2}S_M\frac{1}{2}^+\\\Omega^{4}D_S\frac{1}{2}^+
 \end{array}$\\

$\begin{array}{c} \Omega\frac{3}{2}^+\\\Omega\frac{3}{2}^+ \\
\Omega\frac{3}{2}^+\\\Omega\frac{3}{2}^+
\end{array}$&$
\begin{array}{c} 1673\\ 2078\\ 2208\\ 2263
 \end{array}$&$
\begin{array}{c} 1672\\\\\\\\
 \end{array}$
 &$\left[\begin{array}{rrrr}
  0.99&-0.15& 0.00&0.00\\
 -0.15&-0.98&-0.08&0.04\\
  0.01& 0.08&-0.96&0.27\\
  0.00& 0.02& 0.28&0.96\\
\end{array}\right] $
&$\begin{array}{c}\Omega^{4}S_S\frac{3}{2}^+\\
\Omega
^{4}S'_S\frac{3}{2}^+\\\Omega^{4}D_S\frac{3}{2}^+\\\Omega^{2}D_M\frac{3}{2}^+\end{array}$\\

 $\begin{array}{c}\Omega\frac{5}{2}^+\\\Omega\frac{5}{2}^+
 \end{array}$&$\begin{array}{c}2224\\2260
 \end{array}$&$\begin{array}{c}\\\\
 \end{array}$
 &$\left[\begin{array}{rr}
 0.97& 0.23\\
 - 0.23&0.97\\
 \end{array}\right] $&$\begin{array}{l}\Omega^4D_S\frac{5}{2}^+\\\Omega^2D_M\frac{5}{2}^+
 \end{array}$\\

$\Omega\frac{7}{2}^+$ &2205&&1.00&$\Omega^4D_S\frac{7}{2}^+$\\
\\
$\Omega\frac{1}{2}^-$ &2015&&~1.00&$~\Omega^2P_M\frac{1}{2}^-$\\

$\Omega\frac{3}{2}^-$ &2015&&~1.00&$~\Omega^2P_M\frac{3}{2}^-$\\
\hline\hline

\end{tabular}
\end{lrbox}

\begin{table}[h]
\caption{The calculated  $\Omega$  spectrum  and composition in the
SU(6) basis.  }\label{S-3} \scalebox{1}{\usebox{\tomege}}
\end{table}

\newsavebox{\tlamda}

\begin{lrbox}{\tlamda}
\begin{tabular}{cccll}
   \hline\hline

 State&Mass&Exp.&       \hspace{20mm}Composition                           &\\
    & (MeV)         & (MeV)    &  &\\
 \hline
$\begin{array}{c}\Lambda\frac{1}{2}^+\\\Lambda\frac{1}{2}^+\\\Lambda\frac{1}{2}^+\\\Lambda\frac{1}{2}^+\\
\Lambda\frac{1}{2}^+\\\Lambda\frac{1}{2}^+\\\Lambda\frac{1}{2}^+\\
 \end{array}$&$\begin{array}{c}1113\\1606\\1764\\1880\\2013\\2173\\2198\\
 \end{array}$&$\begin{array}{c}1116\\1560-1700\\1750-1850\\\\\\\\\\
 \end{array}$
 &$\left[\begin{array}{rrrrrrr}
  0.95 & -0.24& -0.17& 0.05& 0.01& 0.00& 0.00\\
  -0.22& -0.95& 0.09 &-0.19& -0.02& 0.00& 0.00\\
  0.12 & 0.18 &0.15  &-0.96& 0.02&0.00&0.00\\
  0.16 & 0.02 & 0.91 &0.16 &-0.34&-0.02&-0.08\\
  0.05 & -0.01& 0.33 &0.08 &0.85 &0.08&0.39\\
  0.01 & 0.00 &0.06  &0.02 & 0.40&-0.28&-0.87\\
  0.00 & 0.00 &0.00  &0.00 & 0.04&0.96&-0.29\\
 \end{array}\right]
 $&$\begin{array}{l}{\Lambda_8}^2S_S\frac{1}{2}^+\\{\Lambda_8}^2S'_S\frac{1}{2}^+\\{\Lambda_8}^2S_M\frac{1}{2}^+\\
 {\Lambda_1}^2S_M\frac{1}{2}^+\\{\Lambda_8}^4D_M\frac{1}{2}^+\\{\Lambda_1}^4P_A \frac{1}{2}^+\\
 {\Lambda_8}^2P_A \frac{1}{2}^+\\
 \end{array}$\\

$\begin{array}{c}\Lambda\frac{3}{2}^+\\\Lambda\frac{3}{2}^+\\\Lambda\frac{3}{2}^+\\\Lambda\frac{3}{2}^+\\
\Lambda\frac{3}{2}^+\\\Lambda\frac{3}{2}^+\\\Lambda\frac{3}{2}^+\\
 \end{array}$&$\begin{array}{c}1836\\1958\\1993\\2061\\2121\\2134\\2174\\
 \end{array}$&$\begin{array}{c}1850-1910\\\\\\\\\\\\\\
 \end{array}$
 &$\left[\begin{array}{rrrrrrr}
  0.11 & -0.80 & -0.34 & 0.47& -0.01& 0.07 & 0.00\\
  0.58 &-0.04  & 0.63  &0.31 & -0.21& -0.34& 0.06\\
  -0.70&-0.28  &0.48   &0.09 & 0.24 &-0.36 &-0.06\\
  0.38 &- 0.10 & 0.00  &-0.21&0.80  &-0.12 &-0.39\\
  -0.12&0.49   &- 0.11 &0.77 &0.19  &0.07  &-0.30\\
  -0.02&-0.14  &0.48   &-0.02& 0.00 &0.83  &-0.24\\
  -0.04&-0.07  &-0.09  &-0.16& -0.47&-0.20 &-0.83\\
 \end{array}\right]
 $&$\begin{array}{l}{\Lambda_8}^4S_M\frac{3}{2}^+\\{\Lambda_8}^2D_S\frac{3}{2}^+\\
 {\Lambda_1}^2D_M\frac{3}{2}^+\\{\Lambda_8} ^2D_S\frac{3}{2}^+\\{\Lambda_8}^4D_M\frac{3}{2}^+\\{\Lambda_1}^4P_A \frac{3}{2}^+\\
 {\Lambda_8}^2P_A \frac{3}{2}^+\\
 \end{array}$\\

$\begin{array}{c}\Lambda\frac{5}{2}^+\\\Lambda\frac{5}{2}^+\\\Lambda\frac{5}{2}^+\\
\Lambda\frac{5}{2}^+\\\Lambda\frac{5}{2}^+\\
 \end{array}$&$\begin{array}{c}1839\\2008\\2103\\2129\\2155\\
 \end{array}$&$\begin{array}{c}1815-1825\\\\\\\\\\
 \end{array}$
 &$\left[\begin{array}{rrrrr}
  -0.82 & -0.33& 0.47  & 0.00 & -0.02\\
  -0.16 & 0.91 & 0.35  &0.06  & 0.13\\
  -0.43 & 0.18 &-0.61  &-0.62 & 0.14\\
  -0.35 & 0.12 & -0.52 &0.73  &-0.24\\
  0.02  & 0.12 & 0.08  &-0.27 &-0.95 \\
 \end{array}\right]
 $&$\begin{array}{l}{\Lambda_8}^2D_S\frac{5}{2}^+\\
 {\Lambda_1}^2D_M\frac{5}{2}^+\\{\Lambda_8}^2D_M\frac{5}{2}^+\\{\Lambda_8}^4D_M \frac{5}{2}^+\\
 {\Lambda_1}^4P_A \frac{5}{2}^+\\
 \end{array}$\\

 $\Lambda\frac{7}{2}^+$ &2064&$2020-2140
$&1.00&${\Lambda_8}^4D_M\frac{7}{2}^+$\\

 \hline\hline
\end{tabular}
\end{lrbox}

\newpage
\begin{table}[h]
\caption{The calculated $\Lambda$ spectrum of positive
parity.}\label{S1lamda} \scalebox{0.75}{\usebox{\tlamda}}
\end{table}

\newsavebox{\tsigma}

\begin{lrbox}{\tsigma}
\begin{tabular}{cccll}
   \hline\hline

  State&Mass&Exp.&       \hspace{20mm}Composition                           &\\
   & (MeV)         & (MeV)    &&\\
 \hline

$\begin{array}{c}\Sigma\frac{1}{2}^+\\\Sigma\frac{1}{2}^+\\\Sigma\frac{1}{2}^+\\
\Sigma\frac{1}{2}^+\\\Sigma\frac{1}{2}^+\\\Sigma\frac{1}{2}^+\\\Sigma\frac{1}{2}^+\\
 \end{array}$&$\begin{array}{c}1192\\1664\\1924\\1986\\2022\\2069\\2172
 \end{array}$&$\begin{array}{c}1193\\1630-1690\\1826-1985\\\\\\\\\\
 \end{array}$
 &$\left[\begin{array}{rrrrrrr}
   0.98 &-0.14&-0.16   & 0.02  & 0.00 & 0.01& 0.00\\
  -0.17 &-0.96 &-0.19   & 0.08  & 0.00 & 0.00& 0.00\\
  -0.12 & 0.17 &-0.86   &-0.20  &-0.18 & 0.36& 0.07\\
   0.04 &-0.08 & 0.35   & 0.03  &-0.81 & 0.44& 0.12\\
  -0.02 & 0.10 &-0.09   & 0.94  &0.09 & 0.28& 0.10\\
  -0.03 & 0.08 &-0.24   & 0.26  &-0.54 &-0.67&-0.36\\
  -0.01 & 0.02 &-0.06   & 0.02  &-0.10 &-0.38& 0.92\\
 \end{array}\right]
 $&$\begin{array}{l}{\Sigma_8}^2S_S\frac{1}{2}^+\\{\Sigma_8}^2S'_S\frac{1}{2}^+\\{\Sigma_8}^2S_M\frac{1}{2}^+\\
 {\Sigma_{10}}^2S_M\frac{1}{2}^+\\{\Sigma_{10}}^4D_S\frac{1}{2}^+\\{\Sigma_8}^4D_M \frac{1}{2}^+\\
 {\Sigma_8}^2P_A \frac{1}{2}^+\\
 \end{array}$\\

$\begin{array}{c}\Sigma\frac{3}{2}^+\\\Sigma\frac{3}{2}^+\\\Sigma\frac{3}{2}^+\\
\Sigma\frac{3}{2}^+\\\Sigma\frac{3}{2}^+\\\Sigma\frac{3}{2}^+\\\Sigma\frac{3}{2}^+\\
\Sigma\frac{3}{2}^+\\\Sigma\frac{3}{2}^+\\
 \end{array}$&$\begin{array}{c}1383\\1868\\1947\\1993\\2039\\2075\\2098\\2122\\2168
 \end{array}$&$\begin{array}{c}1385\\1720-1925\\\\\\\\2070-2140\\\\\\\\
 \end{array}$
 &$\left[\begin{array}{rrrrrrrrr}
   0.98 & 0.17 & 0.01   & 0.00  & 0.00 & 0.00& 0.00& 0.00& 0.00\\
  -0.16 & 0.92 &-0.33   & 0.02  & 0.07 &-0.04& 0.05&-0.04& 0.00\\
  -0.01 & 0.07 & 0.09   &-0.94  &-0.01 & 0.22& 0.07& 0.20&-0.01\\
   0.03 &-0.13 &-0.46   &-0.05  &-0.69 &-0.28& 0.44& 0.12&-0.08\\
   0.05 &-0.26 &-0.62   &-0.20  & 0.54 &-0.43&-0.17&-0.04& 0.04\\
  -0.03 & 0.13 & 0.53   &-0.13  & 0.05 &-0.79& 0.20&-0.12&-0.07\\
  -0.01 & 0.08 & 0.06   & 0.06  &-0.21 &-0.25&-0.55& 0.71& 0.27\\
   0.00 & 0.02 &-0.03   &-0.20  &-0.39 &-0.08&-0.46&-0.64& 0.42\\
   0.00 &-0.02 & 0.03   & 0.06  & 0.18 & 0.05& 0.47& 0.09& 0.86\\
 \end{array}\right]
 $&$\begin{array}{l}{\Sigma_{10}}^4S_S\frac{3}{2}^+\\{\Sigma_{10}}^4S'_S\frac{3}{2}^+\\{\Sigma_8}^4S_M\frac{3}{2}^+\\
{\Sigma_8}^2D_S\frac{3}{2}^+\\
 {\Sigma_{10}}^4D_M\frac{3}{2}^+\\{\Sigma_8}^2D_M\frac{3}{2}^+\\{\Sigma_8}^4D_M \frac{3}{2}^+\\
 {\Sigma_{10}}^2D_M \frac{3}{2}^+\\
 {\Sigma_8}^2P_A \frac{3}{2}^+\\
 \end{array}$\\

$\begin{array}{c}\Sigma\frac{5}{2}^+\\\Sigma\frac{5}{2}^+\\\Sigma\frac{5}{2}^+\\
\Sigma\frac{5}{2}^+\\\Sigma\frac{5}{2}^+\\
 \end{array}$&$\begin{array}{c}1949\\2028\\2062\\2107\\2154\\
 \end{array}$&$\begin{array}{c}1900-1935\\\\2051-2070\\\\\\
 \end{array}$
 &$\left[\begin{array}{rrrrr}
  -0.95 & 0.04 & 0.20   &-0.03 & 0.22\\
  -0.04 &-0.88 & 0.13   &0.44  &-0.08\\
  -0.21 &-0.15 &-0.97   &-0.03 & 0.02\\
   0.21 &-0.04 &-0.02   &0.11  & 0.97\\
  -0.05 & 0.44 &-0.09   &0.89  &-0.08\\
 \end{array}\right]
 $&$\begin{array}{l}{\Sigma_8}^2D_S\frac{5}{2}^+\\
 {\Sigma_{10}}^4D_M\frac{5}{2}^+\\{\Sigma_8}^2D_M\frac{5}{2}^+\\{\Sigma_8}^4D_M \frac{5}{2}^+\\
 {\Sigma_{10}}^2D_M \frac{5}{2}^+\\
 \end{array}$\\

 $\begin{array}{c}\Sigma\frac{7}{2}^+\\\Sigma\frac{7}{2}^+
 \end{array}$&$\begin{array}{c}2002\\2106
 \end{array}$&$\begin{array}{c}2025-2040\\\\
 \end{array}$
 &$\left[\begin{array}{rr}
 0.87& -0.50\\
 0.50& 0.87\\
 \end{array}\right] $&$\begin{array}{r}{\Sigma_{10}}^{4}D_S\frac{7}{2}^+\\{\Sigma_{10}}^{4}D_M\frac{7}{2}^+
 \end{array}$\\
\hline\hline
\end{tabular}
\end{lrbox}

\newpage
\begin{table}[h]
\caption{ The calculated $\Sigma$  spectrum of positive parity.
}\label{S1simga} \scalebox{0.7}{\usebox{\tsigma}}
\end{table}

\newsavebox{\txi}

\begin{lrbox}{\txi}
\begin{tabular}{cccll}
 \hline\hline
  State&Mass&Exp.&       \hspace{20mm}Composition                           &\\
   & (MeV)         & (MeV)    & &\\
 \hline
$\begin{array}{c}\Xi\frac{1}{2}^+\\\Xi\frac{1}{2}^+\\\Xi\frac{1}{2}^+\\
\Xi\frac{1}{2}^+\\\Xi\frac{1}{2}^+\\\Xi\frac{1}{2}^+\\\Xi\frac{1}{2}^+\\
 \end{array}$&$\begin{array}{c}1317\\1750\\1982\\2054\\2107\\2149\\2254
 \end{array}$&$\begin{array}{c}1318\\\\\\\\\\\\\\
 \end{array}$
 &$\left[\begin{array}{rrrrrrr}
  -0.96 & 0.22 & 0.17   & 0.02  & 0.00 &-0.01& 0.00\\
   0.21 & 0.97 &-0.08   & 0.09  & 0.00 & 0.03& 0.00\\
  -0.17 &-0.05 & 0.88   & 0.03  & 0.18 & 0.38& 0.08\\
  -0.06 & 0.13 &-0.38   &-0.11  &-0.72 &-0.54&-0.16\\
  -0.01 &-0.08 &-0.05   & 0.97  & 0.05 &-0.21&-0.09\\
   0.02 &-0.03 & 0.16   & 0.20  &-0.65 & 0.59& 0.38\\
  -0.01 & 0.01 &-0.06   &-0.01  & 0.14 &-0.41& 0.90\\
 \end{array}\right]
 $&$\begin{array}{l}{\Xi_8}^2S_S\frac{1}{2}^+\\{\Xi_8}^2S'_S\frac{1}{2}^+\\{\Xi_8}^2S_M\frac{1}{2}^+\\
 {\Xi_{10}}^2S_M\frac{1}{2}^+\\{\Xi_{10}}^4D_S\frac{1}{2}^+\\{\Xi_8}^4D_M \frac{1}{2}^+\\
 {\Xi_8}^2P_A \frac{1}{2}^+\\
 \end{array}$\\

$\begin{array}{c} \Xi\frac{3}{2}^+\\\Xi\frac{3}{2}^+\\\Xi\frac{3}{2}^+\\
\Xi\frac{3}{2}^+\\\Xi\frac{3}{2}^+\\\Xi\frac{3}{2}^+\\\Xi\frac{3}{2}^+\\
\Xi\frac{3}{2}^+\\\Xi\frac{3}{2}^+\\
 \end{array}$&$\begin{array}{c}1526\\1952\\1970\\2065\\2114\\2174\\2184\\2218\\2252
 \end{array}$&$\begin{array}{c}1530\\\\\\\\\\\\\\\\\\
 \end{array}$
 &$\left[\begin{array}{rrrrrrrrr}
  -0.98 &-0.17 & 0.00   & 0.00  & 0.00 & 0.00& 0.00& 0.00& 0.00\\
  -0.07 & 0.40 & 0.28   &-0.71  & 0.02 & 0.47&-0.03&-0.14& 0.00\\
   0.14 &-0.82 &-0.27   &-0.42  &-0.07 & 0.23& 0.01&-0.04& 0.00\\
   0.03 &-0.14 & 0.44   & 0.15  &-0.63 & 0.13&-0.56& 0.15& 0.11\\
   0.05 &-0.33 & 0.74   & 0.16  & 0.54 & 0.08& 0.12& 0.00&-0.03\\
   0.01 &-0.05 & 0.06   &-0.09  & 0.05 &-0.40&-0.31&-0.83& 0.19\\
  -0.01 & 0.05 &-0.32   & 0.25  & 0.46 & 0.45&-0.52& 0.04& 0.38\\
   0.00 & 0.01 &-0.07   & 0.42  &-0.18 & 0.54& 0.17&-0.49&-0.47\\
   0.00 &-0.01 & 0.07   & 0.14  &-0.24 & 0.18& 0.53&-0.14& 0.76\\
 \end{array}\right]
 $&$\begin{array}{l}{\Xi_{10}}^4S_S\frac{3}{2}^+\\{\Xi_{10}}^4S'_S\frac{3}{2}^+\\{\Xi_8}^4S_M\frac{3}{2}^+\\
{\Xi_8}^2D_S\frac{3}{2}^+\\
 {\Xi_{10}}^4D_M\frac{3}{2}^+\\{\Xi_8}^2D_M\frac{3}{2}^+\\{\Xi_8}^4D_M \frac{3}{2}^+\\
 {\Xi_{10}}^2D_M \frac{3}{2}^+\\
 {\Xi_8}^2P_A \frac{3}{2}^+\\
 \end{array}$\\

 $\begin{array}{c}\Xi\frac{5}{2}^+\\\Xi\frac{5}{2}^+\\\Xi\frac{5}{2}^+\\
\Xi\frac{5}{2}^+\\\Xi\frac{5}{2}^+\\
 \end{array}$&$\begin{array}{c}1959\\2102\\2170\\2205\\2239\\
 \end{array}$&$\begin{array}{c}\\\\\\\\\\
 \end{array}$
 &$\left[\begin{array}{rrrrr}
   0.84 & 0.00 &-0.52   &0.00  & 0.15\\
   0.03 & 0.86 & 0.08   &0.49  & 0.12\\
  -0.21 & 0.14 &-0.57   &0.04  &-0.78\\
   0.46 &-0.18 & 0.58   &0.32  &-0.57\\
  -0.19 &-0.46 &-0.25   &0.81  & 0.19\\
 \end{array}\right]
 $&$\begin{array}{l}{\Xi_8}^2D_S\frac{5}{2}^+\\
 {\Xi_{10}}^4D_M\frac{5}{2}^+\\{\Xi_8}^2D_M\frac{5}{2}^+\\{\Xi_8}^4D_M \frac{5}{2}^+\\
 {\Xi_{10}}^2D_M \frac{5}{2}^+\\
 \end{array}$\\
 $\begin{array}{c}\Xi\frac{7}{2}^+\\\Xi\frac{7}{2}^+
 \end{array}$&$\begin{array}{c}2074\\2189
 \end{array}$&$\begin{array}{c}\\
 \end{array}$&$\left[\begin{array}{rr}
 0.81& 0.58\\
 0.58&-0.81\\
 \end{array}\right] $&$\begin{array}{r}{\Xi_{10}}^{4}D_S\frac{7}{2}^+\\{\Xi_{10}}^{4}D_M\frac{7}{2}^+
 \end{array}$\\

 \hline\hline

\end{tabular}
\end{lrbox}

\begin{table}[h]
\caption{The calculated $\Xi$ spectrum of positive parity.
}\label{S-2} \scalebox{0.7}{\usebox{\txi}}
\end{table}

\newsavebox{\negative}

\begin{lrbox}{\negative}
\begin{tabular}{cccll}
 \hline\hline
  State&Mass&Exp. \hspace{20mm}&       \hspace{20mm}Composition                           &\\
    & (MeV)         & (MeV)    &  &\\
 \hline

$\begin{array}{c} \Lambda\frac{1}{2}^-\\
\Lambda\frac{1}{2}^-\\\Lambda\frac{1}{2}^-
 \end{array}$&$\begin{array}{c}1559 ^a
\\1656\\1791
 \end{array}$&
 $\begin{array}{c}1402-1410\\1660-1680\\1720-1850
 \end{array}$
 &$\left[\begin{array}{rrr}
 0.09& 0.50&0.86\\
 0.59& 0.66&-0.45\\
 0.79& -0.56&0.24\\
 \end{array}\right] $&
 $\begin{array}{c}{\Lambda_8}^{4}P_M\frac{1}{2}^-\\{\Lambda_8}^{2}P_M\frac{1}{2}^-
 \\{\Lambda_1}^{2}P_M\frac{1}{2}^-
 \end{array}$\\

 $\begin{array}{c}
\Lambda\frac{3}{2}^-\\\Lambda\frac{3}{2}^-\\\Lambda\frac{3}{2}^-
 \end{array}$&$\begin{array}{c}1560^a\\1702\\1859
 \end{array}$&
 $\begin{array}{c}1520\\1685-1695\\\\
 \end{array}$
 &$\left[\begin{array}{rrr}
 -0.02& 0.47&0.88\\
 0.13& -0.87&0.47\\
 0.99& 0.12&-0.05\\
 \end{array}\right] $&
 $\begin{array}{c}{\Lambda_8}^{4}P_M\frac{3}{2}^-\\{\Lambda_8}^{2}P_M\frac{3}{2}^-
 \\{\Lambda_1}^{2}P_M\frac{3}{2}^-
 \end{array}$\\
$\Lambda\frac{5}{2}^-$ &1803&$1810-1830$&\ 1.00&${\Lambda_8}^{4}P_M\frac{5}{2}^-$\\

$\begin{array}{c}
\Sigma\frac{1}{2}^-\\\Sigma\frac{1}{2}^-\\\Sigma\frac{1}{2}^-
 \end{array}$&$\begin{array}{c}1657\\1746\\1802
 \end{array}$&$\begin{array}{c}1620\\1730-1800\\\\
 \end{array}$&$\left[\begin{array}{rrr}
0.69& -0.71&-0.15\\
 0.70& 0.60&0.38\\
 -0.18& -0.37&0.91\\
 \end{array}\right] $&
 $\begin{array}{c}{\Sigma_8}^{4}P_M\frac{1}{2}^-\\{\Sigma_8}^{2}P_M\frac{1}{2}^-
 \\{\Sigma_{10}}^{2}P_M\frac{1}{2}^-
 \end{array}$\\
$\begin{array}{c}
\Sigma\frac{3}{2}^-\\\Sigma\frac{3}{2}^-\\\Sigma\frac{3}{2}^-
 \end{array}$&$\begin{array}{c}1698\\1790\\1802
 \end{array}$&$\begin{array}{c}1165-1685\\\\\\
 \end{array}$&$\left[\begin{array}{rrr}
-0.15& -0.95&-0.29\\
 0.89& -0.01&-0.45\\
0.42& -0.33&0.85\\
 \end{array}\right] $&
 $\begin{array}{c}{\Sigma_8}^{4}P_M\frac{3}{2}^-\\{\Sigma_8}^{2}P_M\frac{3}{2}^-
 \\{\Sigma_{10}}^{2}P_M\frac{3}{2}^-
 \end{array}$\\
 $\Sigma\frac{5}{2}^-$ &1743&$1770-1780
 $&~1.00&$~{\Sigma_8}^{4}P_M\frac{5}{2}^-$\\

$\begin{array}{c}
\Xi\frac{1}{2}^-\\\Xi\frac{1}{2}^-\\\Xi\frac{1}{2}^-
 \end{array}$&$\begin{array}{c}1772\\1894\\1926
 \end{array}$&&$\left[\begin{array}{rrr}
0.48& 0.85&-0.19\\
 0.75& -0.29&0.59\\
 0.44& -0.43&-0.78\\
 \end{array}\right] $&
 $\begin{array}{c}{\Xi_8}^{4}P_M\frac{1}{2}^-\\{\Xi_8}^{2}P_M\frac{1}{2}^-
 \\{\Xi_{10}}^{2}P_M\frac{1}{2}^-
 \end{array}$\\
$\begin{array}{c}
\Xi\frac{3}{2}^-\\\Xi\frac{3}{2}^-\\\Xi\frac{3}{2}^-
 \end{array}$&$\begin{array}{c}1801\\1918\\1976
 \end{array}$&$\begin{array}{c}1817-1828\\\\\\
 \end{array}$&$\left[\begin{array}{rrr}
-0.1& 0.95&-0.28\\
 -0.09& 0.27&0.96\\
0.99& 0.12&0.06\\
 \end{array}\right] $&
 $\begin{array}{c}{\Xi_8}^{4}P_M\frac{3}{2}^-\\{\Xi_8}^{2}P_M\frac{3}{2}^-
 \\{\Xi_{10}}^{2}P_M\frac{3}{2}^-
 \end{array}$\\
 $\Xi\frac{5}{2}^-$ &1917&&~1.00&$~{\Xi_8}^{4}P_M\frac{5}{2}^-$\\

\hline\hline

&&$^a$ These results do not& consider the second-order effects in
Ref.~\cite{Isgur and Karl4187}
\end{tabular}
\end{lrbox}

\begin{table}[h]
\caption{The calculated spectrum of the negative-parity S=-1 and
S=-2 excited baryons.
 }\label{negative}
 \begin{center}
\scalebox{0.7}{\usebox{\negative}} \end{center}
\end{table}

\newpage
\newsavebox{\contact}

\begin{lrbox}{\contact}
 \begin{tabular}{lll}
\hline \hline
 $\begin{array}{c} {N_8}^2S_S\frac{1}{2}^+\\{N_8}^2S'_S\frac{1}{2}^+\\ {N_8}^2S_M\frac{1}{2}^+\\
 {N_8}^4D_M\frac{1}{2}^+\\{N_8}^2P_A\frac{1}{2}^+\end{array}$& $\left(\begin{array}{rrrrr}
  0.0293 &-0.014  &-0.0020  & 0.0000&0.0000\\
 -0.0014 & 0.0057 &-0.0334  & 0.0000&0.0000\\
 -0.0020 &-0.0334 & 0.0183  & 0.0000&0.0000\\
  0.0000 &0.0000  & 0.0000  &-0.0259&0.0000\\
  0.0000 &0.0000  & 0.0000  &0.0000 &0.0309\\
 \end{array}\right) $
\hspace{20mm}$\begin{array}{c}{\Delta_{10}}^{4}S_S\frac{3}{2}^+\\{\Delta_{10}}^{4}S'_S\frac{3}{2}^+\\
 {\Delta_{10}}^{4}D_S\frac{3}{2}^+\\{\Delta_{10}}^2D_M\frac{3}{2}^+\end{array}$ $\left(\begin{array}{rrrr}
 -0.0293 & 0.0020 & 0.0000  & 0.0000\\
  0.0020 &-0.0057 & 0.0000  & 0.0000\\
  0.0000 & 0.0000 &-0.0208  & 0.0000\\
  0.0000 &0.0000  & 0.0000  &-0.0259\\
 \end{array}\right) $\\

$\begin{array}{c}{\Lambda_8}^2S_S\frac{1}{2}^+\\{\Lambda_8}^2S'_S\frac{1}{2}^+\\{\Lambda_8}^2S_M\frac{1}{2}^+\\
 {\Lambda_1}^2S_M\frac{1}{2}^+\\{\Lambda_8}^4D_M\frac{1}{2}^+\\{\Lambda_1}^4P_A \frac{1}{2}^+\\
 {\Lambda_8}^2P_A \frac{1}{2}^+\\
 \end{array}$& $\left(\begin{array}{ccccccc}
  0.0293         & 0.0014           & 0.0010-0.0059x   &-0.0010-0.0059x  & 0.0000           & 0.0000           & 0.0000\\
  0.0014         & 0.0057           &-0.0167-0.0147x   & 0.0167-0.0147x  & 0.0000           & 0.0000           & 0.0000\\
  0.0010-0.0059x &-0.0167-0.0147x   & 0.0028+0.0206x   &-0.0028+0.0150x  & 0.0000           & 0.0000           & 0.0000 \\
 -0.0010-0.0059x & 0.0167-0.0147x   &-0.0028+0.0150x   & 0.00284+0.0094x & 0.0000           & 0.0000           & 0.0000\\
  0.0000         & 0.0000           & 0.0000           & 0.0000          &-0.0103-0.0184x   & 0.0000           & 0.0000 \\
  0.0000         & 0.0000           & 0.0000           & 0.0000          & 0.0000           &-0.0103-0.0206x   & 0.0000\\
  0.0000         & 0.0000           & 0.0000           & 0.0000          & 0.0000           & 0.0000           &-0.0103+0.0413x\\
 \end{array}\right) $\\

 $\begin{array}{c}
 {\Sigma_8}^2S_S\frac{1}{2}^+\\{\Sigma_8}^2S'_S\frac{1}{2}^+\\{\Sigma_8}^2S_M\frac{1}{2}^+\\
 {\Sigma_{10}}^2S_M\frac{1}{2}^+\\{\Sigma_{10}}^4D_S\frac{1}{2}^+\\{\Sigma_8}^4D_M \frac{1}{2}^+\\
 {\Sigma_8}^2P_A \frac{1}{2}^+\\
 \end{array}$& $\left(\begin{array}{ccccccc}
 -0.0098+0.0541\hat{x}  & 0.0005-0.0111x  & 0.0003-0.0098x  &-0.0003-0.0020x  & 0.0000             & 0.0000          & 0.0000 \\
  0.0005-0.0111x  &-0.0019+0.0263x  &-0.0056-0.0245x  & 0.0056-0.0049x  & 0.0000             & 0.0000          & 0.0000 \\
  0.0003-0.0098x  &-0.0056-0.0245x  & 0.0145+0.0132x  & 0.0164-0.0188x  & 0.0000             & 0.0000          & 0.0000 \\
 -0.0003-0.0020x  & 0.0056-0.0049x  & 0.0164-0.0188x  & 0.0145+0.0244x  & 0.0000             & 0.0000          & 0.0000 \\
  0.0000          & 0.0000          & 0.0000          & 0.0000          &-0.0069-0.0176x     & 0.0028-0.0047x  & 0.0000 \\
  0.0000          & 0.0000          & 0.0000          & 0.0000          & 0.0028-0.0047x     &-0.0069-0.0199x  & 0.0000 \\
  0.0000          & 0.0000          & 0.0000          & 0.0000          & 0.0000             &0                & 0.0309\\
 \end{array}\right) $\\

$\begin{array}{c}\footnotesize{\Sigma_{10}}^4S_S\frac{3}{2}^+\\{\Sigma_{10}}^4S'_S\frac{3}{2}^+\\{\Sigma_8}^4S_M\frac{3}{2}^+\\
{\Sigma_8}^2D_S\frac{3}{2}^+\\
 {\Sigma_{10}}^4D_M\frac{3}{2}^+\\{\Sigma_8}^2D_M\frac{3}{2}^+\\{\Sigma_8}^4D_M \frac{3}{2}^+\\
 {\Sigma_{10}}^2D_M \frac{3}{2}^+\\
 {\Sigma_8}^2P_A \frac{3}{2}^+\\
 \end{array}$& $\footnotesize\left(\begin{array}{ccccccccc}
 -0.0098-0.0270\hat{x}  &-0.0005+0.0056x  &-0.0005-0.0028x     & 0.0000            & 0.0000         & 0.0000          & 0.0000        & 0.0000        & 0.0000  \\
 -0.0005+0.0056x  &-0.0019-0.0132x  & 0.0079-0.0069x     & 0.0000            & 0.0000         & 0.0000          & 0.0000        & 0.0000        & 0.0000  \\
 -0.0005-0.0028x  & 0.0079-0.0069x  &-0.0019-0.0188x     & 0.0000            & 0.0000         & 0.0000          & 0.0000        & 0.0000        & 0.0000  \\
  0.0000          & 0.0000          & 0.0000             &-0.0070+0.0353x    & 0.0000         &-0.0020-0.0166x  & 0.0000        & 0.0020-0.0033x& 0.0000 \\
  0.0000          & 0.0000          & 0.0000             & 0.0000            &-0.0070-0.0176x & 0.0000          & 0.0028-0.0047x& 0.0000        & 0.0000 \\
  0.0000          & 0.0000          & 0.0000             &-0.0020-0.0166x    & 0.0000         & 0.0120+0.0176x  & 0.0000        & 0.0189-0.0199x& 0.0000 \\
  0.0000          & 0.0000          & 0.0000             & 0.0000            & 0.0028-0.0047x & 0.0000          &-0.0070-0.0199x& 0.0000        & 0.0000 \\
  0.0000          & 0.0000          & 0.0000             & 0.0020-0.0033x    & 0.0000         & 0.0189-0.0199x  & 0.0000        & 0.0120+0.0221x& 0.0000 \\
  0.0000          & 0.0000          & 0.0000             & 0.0000            & 0.0000         & 0.0000          & 0.0000        & 0.0000        & 0.0309\\
 \end{array}\right) $\\
 \hline \hline
\end{tabular}

\end{lrbox}

\begin{table}[h]

\rotcaption{\label{table5}Matrix elements
$H_{\mathrm{contact}}(\mathrm{O}(\alpha_s^2))$ (in units of
$\delta$).}
\rotatebox[origin=c]{90}{\scalebox{0.5}{\usebox{\contact}}}

\end{table}

\newpage
\newsavebox{\tensor}

\begin{lrbox}{\tensor}
\begin{tabular}{lll}
\hline\hline
  $\begin{array}{c} {N_8}^2S_S\frac{1}{2}^+\\{N_8}^2S'_S\frac{1}{2}^+\\ {N_8}^2S_M\frac{1}{2}^+\\
 {N_8}^4D_M\frac{1}{2}^+\\{N_8}^2P_A\frac{1}{2}^+\end{array}$&
$\left(\begin{array}{rrrrr}
 0.0000&0.0000 &0.0000 & 0.0000 & 0.0000\\
 0.0000&0.0000 &0.0000 & 0.0180 & 0.0000\\
 0.0000&0.0000 &0.0000 & 0.0008 & 0.0000\\
 0.0000&0.0180 &0.0008 & 0.0187 &-0.0204\\
 0.0000&0.0000 &0.0000 &-0.0204 & 0.0000\\
 \end{array}\right) $
 \hspace{20mm}$\begin{array}{c}{\Delta_{10}}^{4}S_S\frac{3}{2}^+\\{\Delta_{10}}^{4}S'_S\frac{3}{2}^+\\
 {\Delta_{10}}^{4}D_S\frac{3}{2}^+\\{\Delta_{10}}^2D_M\frac{3}{2}^+\end{array}$ $\left(\begin{array}{rrrr}
  0.0000 & 0.0000 & 0.0000  & 0.0000\\
  0.0000 & 0.0000 & 0.0255  &-0.0180\\
  0.0000 & 0.0255 & 0.0000  & 0.0205\\
  0.0000 &-0.0180 & 0.0205  & 0.0000\\
 \end{array}\right) $\\

 $\begin{array}{c}{\Lambda_8}^2S_S\frac{1}{2}^+\\{\Lambda_8}^2S'_S\frac{1}{2}^+\\{\Lambda_8}^2S_M\frac{1}{2}^+\\
 {\Lambda_1}^2S_M\frac{1}{2}^+\\{\Lambda_8}^4D_M\frac{1}{2}^+\\{\Lambda_1}^4P_A \frac{1}{2}^+\\
 {\Lambda_8}^2P_A \frac{1}{2}^+\\
 \end{array}$& $\left(\begin{array}{ccccccc}
 0.0000   &0.0000            &0.0000         & 0.0000             & 0.0000         & 0.0000           & 0.0000\\
 0.0000   &0.0000            &0.0000         & 0.0000             & 0.0153x        & 0.0000           & 0.0000\\
 0.0000   &0.0000            &0.0000         &0.0000              & 0.0090-0.0037x & 0.0000           & 0.0000 \\
 0.0000   &0.0000            &0.0000         &0.0000              & 0.0090-0.0143x & 0.0000           & 0.0000\\
 0.0000   &0.0153x           &0.0090-0.0037x &0.0090-0.0143x      &-0.0100-0.0270x &-0.0096+0.0151x   &-0.0136-0.0107x\\
 0.0000   &0.0000            &0.0000         &0.0000              &-0.0096+0.0151x & 0.0072+0.0226x   &-0.0101+0.0160x\\
 0.0000   &0.0000            &0.0000         &0.0000              &-0.0136-0.0107x &-0.0101+0.0160x   & 0.0000\\
 \end{array}\right) $\\

$\begin{array}{c}{\Sigma_8}^2S_S\frac{1}{2}^+\\{\Sigma_8}^2S'_S\frac{1}{2}^+\\{\Sigma_8}^2S_M\frac{1}{2}^+\\
 {\Sigma_{10}}^2S_M\frac{1}{2}^+\\{\Sigma_{10}}^4D_S\frac{1}{2}^+\\{\Sigma_8}^4D_M \frac{1}{2}^+\\
 {\Sigma_8}^2P_A \frac{1}{2}^+\\
 \end{array}$& $\left(\begin{array}{rrrrrrr}
0.0000    & 0.0000            & 0.0000          &0.0000          & 0.0000            & 0.0000         & 0.0000\\
0.0000    & 0.0000            & 0.0000          &0.0000          & 0.0120-0.0102x    & 0.0120+0.0051x & 0.0000\\
0.0000    & 0.0000            & 0.0000          &0.0000          &-0.0085+0.0072x    &-0.0085+0.0178x & 0.0000 \\
0.0000    & 0.0000            & 0.0000          &0.0000          & 0.0085+0.0144x    & 0.0085-0.0072x & 0.0000\\
0.0000    & 0.0120-0.0102x    &-0.0085+0.0072x  &0.0085+0.0144x  &-0.0097-0.0168x    &-0.0097+0.0113x & 0.0000\\
0.0000    & 0.0120+0.0051x    &-0.0085+0.0178x  &0.0085-0.0072x  &-0.0097+0.0113x    &-0.0097-0.0352x &-0.0321x \\
0.0000    & 0.0000            & 0.0000          &0.0000          & 0.0000            &-0.0321x        & 0.0000\\
 \end{array}\right) $\\

$\begin{array}{c}\footnotesize{\Sigma_{10}}^4S_S\frac{3}{2}^+\\{\Sigma_{10}}^4S'_S\frac{3}{2}^+\\{\Sigma_8}^4S_M\frac{3}{2}^+\\
{\Sigma_8}^2D_S\frac{3}{2}^+\\
 {\Sigma_{10}}^4D_M\frac{3}{2}^+\\{\Sigma_8}^2D_M\frac{3}{2}^+\\{\Sigma_8}^4D_M \frac{3}{2}^+\\
 {\Sigma_{10}}^2D_M \frac{3}{2}^+\\
 {\Sigma_8}^2P_A \frac{3}{2}^+\\
 \end{array}$& $\footnotesize\left(\begin{array}{ccccccccc}
 0.0000 & 0.0000              & 0.0000             & 0.0000           & 0.0000          & 0.0000           & 0.0000          & 0.0000         & 0.0000\\
 0.0000 & 0.0000              & 0.0000             &-0.0085+0.0072x   &-0.0170+0.0144x  & 0.0060-0.0051x   &-0.0170-0.0072x  &-0.0060-0.0102x & 0.0000\\
 0.0000 & 0.0000              & 0.0000             &-0.0085-0.0036x   &-0.0170-0.0072x  & 0.0060-0.0126x   &-0.0170+0.0250x  &-0.0060+0.0051x & 0.00000\\
 0.0000 &-0.0085+0.0072x      &-0.0085-0.0036x     & 0.0000           & 0.0097-0.0113x  & 0.0000           & 0.0097+0.0057x  & 0.0000         & 0.0000\\
 0.0000 &-0.0170+0.0144x      &-0.0170-0.0072x     & 0.0097-0.0113x   & 0.0000          &-0.0068+0.0080x   & 0.0000          & 0.0068+0.0160x & 0.0000\\
 0.0000 & 0.0060-0.0051x      & 0.0060-0.0126x     & 0.0000           &-0.0068+0.0080x  & 0.0000           &-0.0068+0.0127x  & 0.0000         & 0.0000\\
 0.0000 &-0.0170-0.0072x      &-0.0170+0.0250x     & 0.0097+0.0057x   & 0.0000          &-0.0068+0.0127x   & 0.0000          & 0.0068-0.0080x & 0.0000\\
 0.0000 &-0.0060-0.0102x      &-0.0060+0.0051x     & 0.0000           & 0.0068+0.0160x  & 0.0000           & 0.0068-0.0080x  & 0.0000         & 0.0000\\
 0.0000 & 0.0000              & 0.0000             & 0.0000           & 0.0000          & 0.0000           & 0.0000          & 0.0000         & 0.0000\\
 \end{array}\right) $\\
\hline\hline
\end{tabular}

\end{lrbox}

\begin{table}[h]
\rotcaption{ \label{table6} Matrix elements
$H_{\mathrm{tensor}}(\mathrm{O}(\alpha_s^2))$ (in units of
$\delta$).}
\rotatebox[origin=c]{90}{\scalebox{0.5}{\usebox{\tensor}}}
\end{table}


\begin{thebibliography}{00}

\bibitem {Gursey}
F. G\"{u}rsey and L. A. Radicati, \Journal{\PRL}{13}{173}{1964}.
\bibitem{Sakita}
B. Sakita, \Journal{\PR}{136}{B1756}{1964}.

\bibitem{RujulaCapstick}
A. De R\'{u}jula, H. Georgi, and S. L.~Glashow,
\Journal{\PRD}{12}{147}{1975}.
\bibitem {Capstick}
For a review, see, i.e., S. Capstick and W. Roberts,
\Journal{\PRN}{45}{241}{2000}.

\bibitem {Chodosa}
A. Chodos, R.L. Jaffe, K. Johnson, C.B. Thorn, and V.F. Weisskopf,
\Journal{\PRD}{9}{3471}{1974}.
\bibitem {Chodosb}
A. Chodos, R.L. Jaffe, K. Johnson, and C.B. Thorn,
\Journal{\PRD}{10}{2599}{1974}.
\bibitem {DeGrand}
T. DeGrand, R.L. Jaffe, K. Johnson, and J. Kiskis,
\Journal{\PRD}{12}{2060}{1975}.

\bibitem {Hayasgi140}
A. Hayashi and G. Holzwarth, \Journal{\PLB}{140}{175}{1984}.
\bibitem {M.P.Mattis}
M.P. Mattis and M. Karliner, \Journal{\PRD}{31}{2833}{1985}.

\bibitem {Hooft}
G. 't Hooft, \Journal{\NPB}{72}{461}{1974}.
\bibitem {witten}
E. Witten, \Journal{\NPB}{160}{57}{1979}.

\bibitem {Jenkins}
E. Jenkins, \Journal{\Ann}{48}{81}{1998}


\bibitem {Carlson}
C.E. Carlson, C.D. Carone, J.L. Goity and R.F. Lebed,
\Journal{\PRD}{59}{114008}{1999}.

\bibitem {Matagne}
N. Matagne and Fl. Stancu, \Journal{\NPA}{811}{291}{2008}.

\bibitem {Goity}
J.L. Goity, C. Schat and N.N. Scoccola,
\Journal{\PLB}{564}{83}{2003};

\bibitem {Matagne2005}
N. Matagne and Fl. Stancu, \Journal{\PLB}{631}{7}{2005}.

\bibitem {Allton}
C.R. Allton \textit{et al}. [UKQCD
Collab.],\Journal{\PRD}{60}{034507}{1999}.

\bibitem {Aoki}
S. Aoki \textit{et al.} [CP-PACS Collab.],
\Journal{\PRD}{67}{034503}{2003}.

\bibitem {Aubin}
 C. Aubin \textit{et al.},
\Journal{\PRD}{70}{094505}{2004}.

\bibitem {Jansen}
 K. Jansen, PoS
 \textbf{LAT2008} (2008) 010 arXiv:0810.5634 [hep-lat].

\bibitem {Durr}
S. Durr \textit{et al.},
 \Journal{\Science}{322}{1224}{2008}

\bibitem{Gupta and Radford24}
S.N. Gupta and S.F. Radford,   \Journal{\PRD} {24} {2309}{1981}.


\bibitem{Gupta and Radford25}
S.N. Gupta and S.F. Radford, \Journal{\PRD}{25} {3430}{1982}.

\bibitem{Gupta and Radford26}
S.N. Gupta, S.F. Radford, and W.W. Repko, \Journal{\PRD}
{26}{3305}{1982}.

\bibitem{Titard94}
S. Titard and F.J. Yndur$\mathrm{\acute{a}}$in,
\Journal{\PRD}{49}{6007}{1994};
 \Journal{\PRD}{51}{6348}{1995}

\bibitem{J.Eiglsperger}
J. Eiglsperger, arXiv:0707.1269 [hep-ph] .

\bibitem {Isgur and Karl77}
N. Isgur and G. Karl, \Journal{\PL}{72B}{109}{1977}.

\bibitem {Isgur and Karl78}
N. Isgur and G. Karl, \Journal{\PL}{74B}{353}{1978}.

\bibitem {Isgur and Karl4187}
N. Isgur and G. Karl, \Journal{\PRD}{18}{4187}{1978}.

\bibitem {Isgur and Karl2653}
N. Isgur and G. Karl, \Journal{\PRD}{19}{2653}{1979}.

\bibitem {Isgur and Karl1191}
N. Isgur and G. Karl, \Journal{\PRD}{20}{1191}{1979}.

\bibitem {Kuang-Ta Chao and Isgur 155}
K.-T. Chao, N. Isgur, and G. Karl, \Journal{\PRD}{23}{155}{1981}.

\bibitem{Buchmann97}
A.J. Buchmann, E. Hernandez and A. Faessler,
\Journal{\PRC}{55}{448}{1997}.



\bibitem{Gupta and Radford2690}
S.N. Gupta and S.F. Radford, \Journal{\PRD}{25} {2690}{1982}.

\bibitem{PDG}
Particle Data Group, C. Amsler,  {\it et al.},
\Journal{\PLB}{667}{1}{2008}.

\bibitem {Capstick and Isgur}
S. Capstick and N. Isgur, \Journal{\PRD}{34}{2809}{1986}.

\bibitem {Loring}
U. L$\mathrm{\ddot{o}}$ring, B.C. Metsch  and H.R. Petry,   \Journal{\EPJA}
{10}{395}{2001}

\bibitem {Klempt}
For a review, see, i.e., E. Klempt and J.M. Richard, arXiv:
0901.2055 [hep-ph].

\bibitem {LiuZou}
B.C. Liu, B.S. Zou, \Journal{\PRL} {96} {042002} {2006}

\bibitem {Zou}
B.S. Zou, \Journal{\EPJA} {35} {325} {2008}

\bibitem {Jido}
D. Jido, J.A. Oller, E. Oset, A. Ramos and U.G. Meissner,
\Journal{\NPA} {725} {181} {2003}

\bibitem{Brinkand}
D.M. Brink and G.R. Satchler, {\it Angular Monmentum} (Oxford Univ.
Press, London, England, 1962).
\bibitem{Varshalovich}
 D.A. Varshalovich, A.N.
Moskalev, and V.K. Khersonskii, {\it Quantum Theory of Angular
Momentum} (World Scientific, Singapore, 1988)

\bibitem{A.C.Aguilar}
A.C. Aguilar, A. Mihara, and A.A. Natale,
\Journal{\PRD}{65}{054011}{2002}.
\end{thebibliography}
\end{document}